\documentclass[11pt,a4paper]{article}
\usepackage{jcappub}
\usepackage{bm}
\usepackage{color}

\newcommand{\Mpl}{M_{\text{pl}}}
\newcommand{\N}{\mathcal{N}}

\newcommand{\G}{G}
\newcommand{\F}{\mathcal{F}}
\newcommand{\M}{\mathcal{M}}
\renewcommand{\P}{\mathcal{P}}

\newcommand{\vp}{\varphi}

\newcommand{\grad}{\nabla}
\newcommand{\im}{\mathrm{i}}

\newcommand{\timeorder}{\mathsf{T}}
\newcommand{\pathorder}{\mathsf{P}}
\newcommand{\e}[1]{\mathrm{e}^{{#1}}}
\newcommand{\EulerGamma}{\gamma_{\mathrm{E}}}

\renewcommand{\d}{\mathrm{d}}

\newcommand{\bigamma}{T}
\newcommand{\biN}{N}

\newcommand{\Dl}{\mathcal{D}_\lambda}
\newcommand{\Dt}{\mathcal{D}_t}
\newcommand{\Deta}{\mathcal{D}_\eta}
\newcommand{\DN}{\mathcal{D}_N}

\newcommand{\trajp}{\Pi}

\newcommand{\vect}[1]{\bm{\mathrm{{#1}}}}
\newcommand{\fnl}{f_{\mathrm{NL}}}

\newcommand{\be}{\begin{equation}}
\newcommand{\ee}{\end{equation}}
\newcommand{\bea}{\begin{eqnarray}}
\newcommand{\eea}{\end{eqnarray}}

\renewcommand{\geq}{\geqslant}

\DeclareMathOperator{\Or}{O}

\renewcommand{\baselinestretch}{1.05}
\notoc

\begin{document}

\title{The inflationary bispectrum with curved field-space}

\author{Joseph Elliston$^a$,}
\author{David Seery$^b$}
\author{and Reza Tavakol$^a$.}

\affiliation{$^a$ School of Physics and Astronomy, Queen Mary University of
London, Mile End Road, London, E1 4NS, UK}

\affiliation{$^b$ Astronomy Centre, University of Sussex, Pevensey II
building, Brighton, BN1 9QH, UK}

\emailAdd{j.elliston@qmul.ac.uk}
\emailAdd{d.seery@sussex.ac.uk}
\emailAdd{r.tavakol@qmul.ac.uk}

\abstract{We compute the covariant three-point function 
near horizon-crossing for a system of slowly-rolling scalar fields 
during an inflationary epoch, allowing for an arbitrary field-space metric. 
We show explicitly how to compute its subsequent evolution using a
covariantized version of the separate universe or `$\delta N$' expansion,
which must be augmented by terms measuring curvature of the 
field-space manifold, and give the nonlinear gauge transformation 
to the comoving curvature perturbation.
Nonlinearities induced by the field-space curvature terms
are a new and potentially significant source of non-Gaussianity.
We show how inflationary models with non-minimal coupling
to the spacetime Ricci scalar can be accommodated within this framework.
This yields a simple toolkit allowing
the bispectrum to be computed in
models with non-negligible field-space curvature.}

\keywords{Inflation, non-Gaussianity, non-canonical, bispectrum}

\maketitle
    
\section{Introduction} \label{sec:introduction}

Considerable effort has recently been invested
in the study of multiple-field models of inflation.
There are three principal motivations:
First, unless curvature couplings flatten their potentials at
high energy density,
the Standard Model has not produced scalar fields which can
successfully inflate. This has stimulated the search for 
realizations
of
inflation in theories beyond the Standard Model. 
Second, single-field inflationary models often require super-Planckian
field excursions. Unfortunately, well-rehearsed arguments suggest that we
should not expect the scalar potential to be stable against
renormalization-group running over such large distances in field space.
Multiple-field models may evade this problem by allowing sub-Planckian
excursions.
Finally, interactions between several fields can give rise to observable
non-Gaussianity. This
may make some multiple-field models sufficiently predictive
that they can be falsified by observation.

It has been known for some time that,
under slow-roll conditions,
multiple-field models with canonical kinetic terms
generate unobservable three- and four-point functions at horizon-crossing
\cite{Seery:2005gb,Seery:2006vu,Seery:2008ax,Chen:2010xka}. Observable effects
can arise only later, from nonlinear processes operating on superhorizon
scales.
A reasonably clear picture has emerged in which these nonlinear effects can be
understood as the deformation of a Gaussian probability distribution by the
phase space flow associated with the theory
\cite{Mulryne:2009kh,Mulryne:2010rp,Elliston:2011dr,Seery:2012vj}.

Noncanonical kinetic terms offer new possibilities. In some
theories the Lagrangian becomes an arbitrary function of the kinetic energy
$2X = - \partial_\mu \vp \partial^\mu \vp$. Where this yields a reduced
sound speed for perturbations there
can be a significant enhancement of the three- and four-point functions
\cite{Alishahiha:2004eh,Seery:2005wm,Chen:2006nt,Cheung:2007st,
Arroja:2008yy,Langlois:2008qf,RenauxPetel:2011uk}.
But this is not the only type of noncanonical Lagrangian. In many examples
descending from our present ideas about physics at very high
energies,
including string theory and supergravity, the kinetic energy must be written
$2X = - \G_{IJ} \partial_\mu \vp^I \partial^\mu \vp^J$, where $\G_{IJ}$ is
an arbitrary, symmetric function of the fields $\vp^I$.
The simplest example is the nonlinear $\sigma$-model of
Gell-Mann~\&~L\'{e}vy, originally introduced to describe spin-0 mesons.

The matrix $\G_{IJ}$ can be thought of as a metric on the space of scalar
fields and will generically exhibit nonzero curvature.
Are there interesting enhancements of non-Gaussian effects in these models?
Estimates of the three-point function have been made by a number of
authors~\cite{Sun:2005ig,Langlois:2008qf,RenauxPetel:2011uk,
Peterson:2011yt,Saffin:2012et}, but as yet no complete 
formalism exists which allows the bispectrum to be followed
from horizon-crossing up to the time of observation.
Moreover, Gong \& Tanaka recently pointed out that 
the most widely-used
formulation of nonlinear perturbation
theory is not covariant \cite{Gong:2011uw}.
They introduced a covariant description,
to be discussed in \S\ref{sec:covariant-perts},
and constructed
the action for fluctuations up to third-order
in the scenario of Langlois~et~al.~\cite{Langlois:2008qf}.
A similar argument was later made by Saffin~\cite{Saffin:2012et}.
Covariance is a convenience rather than a physical principle, 
so its absence does not invalidate earlier results.
Nevertheless, it is a considerable convenience:
there are subtleties associated with
time evolution of the two-point function on curved field-space
which are most clearly expressed in covariant language.
These take the form of a contribution to the
effective mass-matrix from the Riemann curvature tensor.
In this paper we show that a similar phenomenon occurs for the
three-point function.

In flat field-space, time evolution 
of superhorizon modes may be taken into 
account using the `separate universe'
method~\cite{Starobinsky:1982ee,Lyth:1984gv,Starobinsky:1986fxa,Lyth:2005fi}.
This enables the time dependence of each fluctuation to be
determined from the relative behaviour of separated spatial
regions following slightly displaced phase-space trajectories.
It can be effected using a Taylor expansion
to compare two solutions of the slow-roll equation.
But in curved field-space we must be cautious when comparing
the relative motion of neighbouring trajectories.
In the analogous case of general relativity
one would use the equation of geodesic deviation, 
or `Jacobi equation'.
In flat field-space this can be integrated to reproduce
the Taylor expansion~\cite{Seery:2012vj}.
When promoted to curved field-space the Jacobi approach is 
automatically covariant and accounts naturally for 
time-dependent effects generated by the Riemann curvature,
including its known contribution to the
effective mass-matrix.
Beyond linear order there are new contributions which influence
the three-point function.
These must appear in any correct formulation but are expressed 
most clearly and economically in terms of the field-space curvature.

At present, the covariant approach cannot be used to
generate observable predictions
beyond the power spectrum.
To do so would require a
determination of the initial value
of the three-point function near horizon-crossing,
together with a prescription to evolve it
into the 
primordial curvature perturbation. It is only the 
curvature perturbation which can be connected with 
observable quantities.
Neither of these pre-requisites has yet been provided.

In this paper we compute the 
initial value of the covariant three-point function
at horizon exit and use the Jacobi approach to determine 
its time evolution.
(Partial expressions for the noncovariant three-point function
were given by Langlois~et~al.~\cite{Langlois:2008qf}
and Renaux-Petel~et~al.~\cite{RenauxPetel:2011uk}.)
As a concrete example we give the analysis
for the $\sigma$-model Lagrangian $\mathcal{L} = X + V$,
although our methods generalize to more complex cases.
We show that this initial value can be smoothly
connected to the subsequent Jacobi evolution.
In particular, the evolutionary effects
described above which depend on the
Riemann tensor can be matched to new infrared divergences
in the three-point function.
We demonstrate this matching explicitly to subleading order in both
time-dependent perturbation theory and the slow-roll expansion.
A further benefit of the Jacobi
approach is that time evolution can be computed very simply
using ordinary differential equations.

In~\S\ref{sec:perturbations} we specialize the results of
Gong~\&~Tanaka to the $\sigma$-model Lagrangian
and obtain the action for fluctuations to third-order. 
In~\S\ref{sec:n-pfs} we compute the corresponding two- and three-point
functions near horizon-crossing in the spatially flat gauge. The two-point
function has been known since the work of
Sasaki~\&~Stewart~\cite{Sasaki:1995aw}, but the computation of the
covariant three-point function is new. 
In~\S\ref{sec:evolution} we use the Jacobi method discussed 
above to compute the evolution of these correlation 
functions after horizon-crossing.

In~\S\ref{sec:action} we show that our results can be applied to models in
which the scalar fields are coupled non-minimally to gravity by making a
suitable conformal transformation \cite{Kaiser:2010ps}. Such couplings arise
naturally in the low-energy limit of higher-dimensional theories including
supergravity, string theory and Kaluza--Klein models
\cite{Rainer:1996gw,Gunther:1997ft,Appelquist:1987nr}, or as counterterms
in curved spacetime \cite{Birrell:1982ix,Buchbinder:1992rb}. Finally,
we conclude in~\S\ref{sec:conclusion}.

\paragraph{Notation.}
Throughout, we work in units where $c = \hbar = 1$ and express the
gravitational coupling in units of the reduced Planck mass,
$\Mpl^{-2} \equiv 8 \pi G$. 
Upper-case Latin indices $I$, $J$, $K$, \ldots, label the species of
scalar fields, and Greek letters label spacetime indices.
We use a modified index convention for bilocal tensors
(`bitensors'), to be described in \S\ref{sec:third_order}.
The covariant derivative compatible with the field-space metric
$\G^{IJ}$ is $\grad_I$. For any tensor $\mathcal{F}_{\cdots}$ we write
$\grad_I \mathcal{F}_{\cdots} = \mathcal{F}_{\cdots;I}$.
Our sign convention for the curvature tensor is
defined by the Ricci identity, $[ \grad_I, \grad_J ] V_K = R_{IJKL} V^L$.
Finally, it is useful to define covariant versions of the derivatives with respect
to coordinate time $t$, conformal time
$\eta = \int \d t / a$ and e-folds
$N = \int H \, \d t$ as
\begin{equation}
	\Dt = \frac{\d \phi^I}{\d t} \grad_I ,
	\qquad
	\Deta = \frac{\d \phi^I}{\d \eta} \grad_I ,
	\qquad
	\DN = \frac{\d \phi^I}{\d N} \grad_I .
\end{equation}

\section{The action and its perturbations} \label{sec:perturbations}

Consider an inflationary epoch driven by $\N$ scalar fields $\vp^I$
(with $I=1,2, \ldots, \N$), minimally coupled to gravity and self-interacting
through a potential $V(\vp)$,
\begin{equation}
	\label{eq:action}
	S = \frac{1}{2} \int \d^4 x \, \sqrt{-g} \left[ \Mpl^2 R -
	\G_{IJ} g^{\mu \nu} \partial_\mu \vp^I \partial_\nu \vp^J - 2 V \right] .
\end{equation}
As explained in \S\ref{sec:introduction},
the field space is to be understood as a manifold with
metric $\G_{IJ}(\vp)$. This metric is used to raise and lower
tangent-space indices, and distinguishes the model from 
canonical scenarios where $\G_{IJ} = \delta_{IJ}$.

During inflation the $\vp^I$ take approximately homogeneous 
but time-dependent values which we label $\phi^I(t)$. 
Small inhomogeneities around this homogeneous background are 
controlled by a perturbative expansion of~\eqref{eq:action}. 
When quantized they will
seed the primordial inflationary fluctuation.

\subsection{Covariant perturbations}
\label{sec:covariant-perts}

In what follows we perturbatively expand the action for these
fluctuations to third-order. On curved field-space it is very helpful
to arrange this expansion so that covariance is manifest. Each
inhomogeneity is a coordinate displacement 
$\delta \vp^I = \vp^I(\vect{x},t) - \phi^I(t)$, but for finite length this
displacement does not lie in the tangent space at $\phi^I$ and therefore
does not have a tensorial transformation law. 
To obtain a covariant description we must find an alternative
characterization which associates each displacement with a tangent-space
vector. This construction was given by Gong \& Tanaka~\cite{Gong:2011uw},
whose method we briefly describe.

We assume the field-space metric to be smoothly differentiable in the
neighbourhood of the background trajectory.
Within a normal neighbourhood,
the points $\vp^I(\vect{x},t)$ and $\phi^I(t)$ are linked
by a unique geodesic, labelled by a parameter $\lambda$. 
We adjust the normalization so that $\lambda = 0$ corresponds to the unperturbed coordinate $\phi^I$ and $\lambda = 1$ corresponds to the perturbed coordinate $\phi^I + \delta \vp^I$. 
The coordinate displacement $\delta \vp^I$
may be expressed as a formal Taylor series along this geodesic
\begin{equation}
	\delta \vp^I
	\equiv 
	\left. \frac{\d \vp^I}{\d \lambda} \right|_{\lambda=0} + 
	\frac{1}{2!} \left. \frac{\d^2 \vp^I}{\d \lambda^2} \right|_{\lambda=0} + 
	\frac{1}{3!} \left. \frac{\d^3 \vp^I}{\d \lambda^3} \right|_{\lambda=0} +
	\cdots .
	\label{eq:perturbation-expansion}
\end{equation}
Eq.~\eqref{eq:perturbation-expansion} is independent of the
normalization of $\lambda$, so our particular choice was 
merely a convenience. The geodesic satisfies
\begin{equation}
	\Dl^2 \vp^I =
	\frac{\d^2 \vp^I}{\d \lambda^2}
	+ \Gamma^I_{JK} \frac{\d \vp^J}{\d \lambda} \frac{\d \vp^K}{\d \lambda}
	= 0 ,
	\label{eq:geodesic}
\end{equation}
where $\Dl \equiv Q^I \grad_I$
and $Q^I \equiv \d \vp^I / \d \lambda |_{\lambda = 0}$.
Using~\eqref{eq:geodesic}, the expansion 
\eqref{eq:perturbation-expansion} can be reorganized 
as a power series in $Q^I$, yielding
\begin{equation}
	\label{eq:Q_expansion}
	\delta\vp^I =
	Q^I
	-\frac{1}{2!} \Gamma^I_{JK} Q^J Q^K
	+ \frac{1}{3!} \left( \Gamma^I_{LM} \Gamma^M_{JK} - \Gamma^I_{JK,L} \right)
		Q^J Q^K Q^L
	+ \cdots ,
\end{equation}
where the coefficients $\Gamma^I_{JK}$,
$\Gamma^I_{JK,L}$, \ldots, are evaluated at $\lambda = 0$.
In flat field-space all terms but the first vanish and therefore
$\delta \vp^I = Q^I$. It was to achieve this correspondence that we adopted
our normalization for $\lambda$. 
Although~\eqref{eq:Q_expansion} is not itself covariant, it can be used to
exchange an expansion in $\delta\vp^I$ for an expansion in $Q^I$.
Since $Q^I$ \emph{does} lie in the tangent space at $\phi^I$,
the perturbative
expansion of any tensorial object will be manifestly covariant if expressed
in powers of $Q^I$.
We label tangent-space indices at the perturbed position
$\phi^I + \delta\vp^I$ by primed indices $I'$, $J'$
and tangent-space indices at the original position
$\phi^I$ by unprimed indices.
For any tensor $\F_{I \cdots J}$ we obtain
\begin{equation}
	\F_{I' \cdots J'} = {\G_{I'}}^I \cdots {\G_{J'}}^J
		\left(
		\left. \F_{I \cdots J} \right|_{\lambda=0}
		+ \left. \Dl \F_{I \cdots J} \right|_{\lambda=0}
		+ \frac{1}{2!} \left. \Dl^2 \F_{I \cdots J} \right|_{\lambda=0}
		+ \cdots
	\right) ,
	\label{eq:covariant-expansion}
\end{equation}
where ${\G_{I'}}^{I}$ is the parallel propagator,
which expresses parallel transport along the geodesic
connecting $\phi^I$ with $\phi^I + \delta\vp^I$.
For details, see Poisson, Pound~\&~Vega~\cite{lrr-2011-7}.

\subsection{Gauge choice and infrared-safe observables}
\label{sec:ir-safety}
Inhomogeneities in $\vp^I$ couple to gravity and therefore induce
fluctuations in the metric. The description of this mixing is simplified
using ADM variables~\cite{Arnowitt:1962hi}, in terms of which the metric
can be written
\begin{equation}
	\label{eq:ADM}
	\d s^2 = - N^2 \, \d t^2
		+ h_{ij}(\d x^i + N^i \,\d t)(\d x^j + N^j \,\d t) ,
\end{equation}
where $N$ is the lapse function and 
$N^i$ is the shift vector.
Spatial indices
are raised and lowered using the 3-metric $h_{ij}$, which has
determinant $h$. In these variables,
the action can be written
\begin{equation}
	S = \frac{1}{2} \int \d^4 x \, \sqrt{h} \left\{
		\Mpl^2
		\left[
			N R^{(3)} + \frac{1}{N} (E_{ij}E^{ij}-E^2)
		\right]
		+ \frac{1}{N} \pi^I \pi_I
		- N \partial_i \vp^I \partial^i \vp_I
		- 2 N V
	\right\} ,
\label{eq:ADM-action}
\end{equation}
where $R^{(3)}$ is the Ricci scalar built from the 3-metric 
and
$E_{ij}$ is proportional to the extrinsic curvature of slices of constant $t$,
\begin{equation}
	2 E_{ij} = \dot h_{ij} - N_{i|j} - N_{j|i} .
\end{equation}
We have defined $E$ to be the trace ${E^i}_i$, and a vertical bar denotes
the covariant derivative compatible with $h_{ij}$.
Finally, $\pi^I = \dot{\vp}^I - N^j {\vp^I}_{|j}$
and an overdot denotes a derivative with respect to $t$.

The lapse and shift appear in~\eqref{eq:ADM-action} without 
time derivatives. Therefore they are not propagating modes, 
but yield constraint equations.
These determine $N$ and $N^i$ as functions of the 
physical degrees of freedom, and also generate gauge 
symmetries associated with translations in time and space. 
After imposing the constraints and removing redundant gauge 
modes we are left with $\N$ scalar degrees of freedom plus two
polarizations of the graviton which transform as a spin-2 mode, 
but we are free to choose how the scalar modes are divided 
between the $\vp^I$ and $h_{ij}$ by making gauge transformations.

\paragraph{Infrared-safe observables.}
When selecting a gauge we must balance competing demands. 
First, consider a single-field model where $\N = 1$. 
A common gauge choice is to arrange slices of constant $t$ 
to coincide with slices of constant $\vp$, 
leaving a scalar metric mode which we label $\zeta$.
This scalar mode is a perturbation to the volume element,
$\zeta = (1/6) \delta \ln \det h$. 
The disadvantage of this choice is that the calculation is long
\cite{Maldacena:2002vr}. Multiple partial integrations
are required to bring the action to a suitable form, 
simultaneously generating boundary terms which must be
retained~\cite{Adshead:2011bw,Arroja:2011yj,Burrage:2011hd}.

The advantage is a precise type of technical simplicity. 
With only a single field it is a classical theorem that $\zeta$ 
is time independent~\cite{Lyth:2004gb,Rigopoulos:2003ak}.
The same is true in the quantum theory, at least at tree-level,
for the correlation functions of $\zeta$ \cite{Weinberg:2008mc}.%
	\footnote{A similar statement can be made at one-loop level, 
	ignoring internal graviton lines~\cite{Senatore:2009cf,Pimentel:2012tw}.	
	Even if present, loop-generated time dependences are 
	typically strongly suppressed~\cite{Burgess:2009bs}. 
	Although they might be important to describe the evolution 
	of correlations over very large time and distance scales, 
	it seems probable they would be negligible for the 
	description of a phase of observable slow-roll inflation.}
If present, time dependence would manifest itself as a divergence in the
vertex integrals appearing in each $n$-point
function~\cite{Zaldarriaga:2003my,Seery:2007wf,Seery:2010kh}. 
Each integral sums the amplitude for an interaction to occur 
at a specific time, and a divergence means only that 
interactions continue arbitrarily far into the future. 
At tree-level the great merit of
$\zeta$-gauge for
single-field models is that the $n$-point functions 
are free of divergences: 
each vertex integral receives significant contributions only from
interactions which occur near horizon-crossing.
Physically, predictions
for $\zeta$ decouple from the infrared dynamics of the theory.

For this reason we describe $\zeta$ as `infrared safe'.%
	\footnote{The terminology is borrowed from gauge theories, where an 
	infrared-safe observable dominated by a hard subprocess	occurring 
	near energy $E$ does not depend on the details of other processes 
	(such as hadronization and confinement)	occurring at energies much 
	less than $E$. A similar discussion was given by 
	Weinberg~\cite{Weinberg:2005vy}, who focused on the conditions
	under which (in our language) observables might be infrared-safe.
	(However, there is no reason of principle for any physical observable 
	of interest	to be infrared-safe.)}
In practice this means that subleading terms in the action
map to subleading terms in each correlation function.
This makes it simple to impose approximation schemes, such as
the slow-roll expansion.

The enormous convenience of infrared safety means that most single-field
calculations have made use of this gauge.
An alternative is to fix
slices of constant $t$ to carry a flat metric, leaving a perturbation in
the scalar field value $\vp(\vect{x},t)$.
In view of the discussion in \S\ref{sec:covariant-perts} we denote the
covariant representation of this perturbation $Q^I$.
Calculation of the lowest-order action for $Q^I$ is simpler than in the
$\zeta$-gauge \cite{Seery:2005gb}.
But unlike $\zeta$ the field perturbation is not infrared safe because
the integrals which define its correlation functions receive contributions
from all times, not just those near horizon
crossing~\cite{Falk:1992sf,Stewart:1993bc,Nakamura:1996da,
Zaldarriaga:2003my,Seery:2008qj}.
Therefore they are sensitive to the infrared dynamics of the theory.
As a result, subleading terms in the action can be
enhanced by divergences and special action is required
to deal with them.
In a single-field model, Maldacena argued that they
can be accounted for by a gauge transformation to $\zeta$ just after horizon crossing~\cite{Maldacena:2002vr}.

\paragraph{Multiple-field models.}
In a multiple-field model the situation is more complex.
Because $\zeta$ is evolving,
the divergences can no longer be accounted for using only a
gauge transformation on a fixed time-slice.
The solution proposed in Ref.~\cite{Seery:2005gb} was to evaluate each
$n$-point function in the spatially flat gauge a few e-folds after
horizon crossing. This choice prevents enhanced subleading terms
from spoiling
the lowest-order slow-roll prediction, and therefore we can take
advantage of the computational simplicity of this gauge. 
The price we pay is that some means must be found to express the
correlation functions at a subsequent time in terms of their values near
horizon crossing. However this is done, the result must equal what would have
been obtained if we had been able to evaluate the original divergent integrals.
Therefore,
in a perturbative expansion,
it must reproduce the same
divergences. For canonical fields on flat field-space, the nonlinear
separate universe or `$\delta N$' formalism introduced by
Lyth~\&~Rodr\'{\i}guez can be used for this purpose~\cite{Lyth:2005fi}.

On curved field-space we expect new divergences involving
the field-space Riemann curvature~\cite{Gong:2011uw},
and we must be aware that these will modify the time-evolution
of the correlation functions.
We will argue in \S\ref{sec:evolution} that these can be understood
as an analogue of the geodesic-deviation effect
for freely-falling observers
in curved spacetime,
and show how they can be incorporated in a new version of the
`$\delta N$' formula.

The conclusion of this discussion is that we are still entitled to choose
the spatially flat gauge in order to simplify the calculation. However,
we must take care to study the effect of divergent terms. 
Although we will carry the QFT calculation of the three-point function only
as far as horizon-crossing, where the divergences are harmless,
it is important to check carefully that whatever technique we
employ to account for time dependence correctly reproduces all these
divergent pieces.

\section{Two- and three-point functions of the fluctuations}
\label{sec:n-pfs}

\paragraph{Flat gauge calculation.}
With this in mind, we specialize to the spatially flat gauge where
$h_{ij} = a^2 \delta_{ij}$. To expand~\eqref{eq:action} to
third-order in $Q^I$ we need only solve the constraints for the
first-order components of $N$ and $N^i$~\cite{Maldacena:2002vr,Chen:2006nt}.
The shift vector can be decomposed into irrotational and solenoidal parts,
yielding
$N^i \equiv \partial^i \vartheta + \beta^i$ where $\partial_i \beta^i=0$. 
The first-order solenoidal component $\beta^i_1$ appears uncoupled to the
other perturbations in the second-order action $S_{(2)}$,
yielding the constraint equation $\beta^i_1=0$.
The remaining metric perturbations can be expanded in powers of $Q$,
\begin{align}
	N &= 1 + \alpha_1 + \alpha_2 + \cdots , \nonumber \\
	\vartheta &= \vartheta_1 + \vartheta_2 + \cdots .
\end{align}
In what follows we determine these metric fields
and use them to obtain the
two- and three-point functions for the $Q^I$.

\subsection{Linear order} \label{sec:first_order}

At linear order we find
\begin{equation}
	S_{(1)}
	= \int \d^4 x \left\{
		a^3 \left[
			3\Mpl^2 H^2
			- \frac{1}{2} \dot \phi_I \dot \phi^I - V
		\right] \alpha_1 
		- \left[
			\Dt (a^3 \dot \phi_I)
			+ a^3 V_{,I} \right]
		Q^I
	\right\} ,
\end{equation}
where we have integrated by parts and removed total derivatives.
The background field equations follow after varying this action with respect to $\alpha_1$ and $Q^I$,
\begin{align}
	3 \Mpl^2 H^2 & =
		\frac{1}{2} \dot \phi_I \dot \phi^I + V
	\label{eq:Friedmann} ,
	\\
	\Dt \dot{\phi}_I + 3H \dot \phi_I
		& = - V_{,I} .
	\label{eq:KG}
\end{align}
The slow-roll regime in curved field-space was discussed by
Sasaki~\&~Stewart~\cite{Sasaki:1995aw} and later
Nakamura~\&~Stewart~\cite{Nakamura:1996da}.
Inflation occurs when $\epsilon \equiv - \dot{H}/H^2 < 1$.
For a successful phenomenology we also require that inflation is
sufficiently prolonged. Therefore $\epsilon$ should not change significantly
in a Hubble time. To satisfy these requirements we must choose
\begin{equation}
	\epsilon =
		\frac{1}{2 \Mpl^2}
		\frac{\dot{\phi}^I \dot{\phi}_I}{H^2} \ll 1
	\qquad \text{and} \qquad
	\eta \equiv \frac{1}{H} \frac{\d \ln \epsilon}{\d t}
		= \frac{2}{H} \frac{\dot{\phi}^I
		\Dt \dot{\phi}_I}{\dot{\phi}^J \dot{\phi}_J}
		+ 2 \epsilon \ll 1 .
\end{equation}
The condition $\eta \ll 1$ requires the tangential component of
the acceleration vector $H^{-1} \Dt \dot{\phi}^I$ (measured in Hubble units)
to be much smaller than the tangent vector to the trajectory.

One can verify that the slow-roll equation
\begin{equation}
	3 H \dot \phi_I + V_{,I} \approx 0
	\label{eq:sr_eom}
\end{equation}
gives a self-consistent 
realization
of these conditions if the potential is
sufficiently flat. Detailed conditions are given in
Refs.~\cite{Sasaki:1995aw,Nakamura:1996da}.
However, global flatness of the potential is not necessary.
This possibility has received recent
attention~\cite{Tolley:2009fg,Achucarro:2010jv,Achucarro:2010da,
Avgoustidis:2012yc}. 
In this paper we do not assume any particular properties of the background
theory, except that it realizes
an era of inflation with $\epsilon \ll 1$
and in which $H$ varies smoothly during horizon exit. 
Such scenarios are not the only cases of interest, but a dedicated analysis
is required where the background evolution exhibits a
feature~\cite{Adshead:2011bw,Adshead:2011jq}.

When we compute the two- and three-point functions we will do so only for
field-space directions which are light during horizon exit.
Therefore our results will not apply to any heavy directions
generated by a steep potential orthogonal to the inflationary trajectory.
(However, when estimating the magnitude of terms in the second-
and third-order actions,
we quote powers of $\dot{\phi}^I/H$ to emphasize that individual components
of this vector are not necessarily of order $\epsilon^{1/2}$.)
This is sufficient to estimate the statistics of the primordial
curvature perturbation in an approximation where the fluctuations in massive
directions at horizon exit are neglected. 
In simple models this is acceptable because large masses rapidly drive any
fluctuations to extinction.
In more complex models it has been suggested that
modest corrections can occur
where the phase-space flow drives power
from massive modes into the curvature perturbation before
decay~\cite{Chen:2009zp,Chen:2012ge,McAllister:2012am}.
To capture these effects would require an extension of the formalism
of~\S\S\ref{sec:second_order}--\ref{sec:third_order} used to compute
initial conditions, although we expect that the subsequent transfer of
power would be correctly described by the method of~\S\ref{sec:evolution}.

\subsection{Second order} \label{sec:second_order}

Expanding the action to second order,
performing multiple partial integrations and removing
total derivatives, we find
\begin{equation}
\begin{split}
	S_{(2)} =
	\frac{1}{2} \int \d^4 x \, a^3 \Big\{
	& \alpha_1 \Big[
		-6 \Mpl^2 H^2 \alpha_1
		+ \dot \phi_I \dot \phi^I \alpha_1
		- 2 \dot \phi_I \Dt Q^I
		- 2  V_{,I}Q^I
	\Big]
	\\
	&
	\mbox{}
	- \frac{2}{a^2} \partial^2 \vartheta_1 \Big[
		2 \Mpl^2 H \alpha_1
		- \dot \phi_I Q^I
	\Big]
	\\
	&
	\mbox{}
	+ R_{KIJL} \dot \phi^K \dot \phi^L Q^I Q^J 
	+ \Dt Q_I \Dt Q^I
	- h^{ij} \partial_i Q_I \partial_j Q^I
	- V_{;IJ}Q^I Q^J
	\Big\} .
	\label{eq:2nd_order}
\end{split}
\end{equation}
The momentum and energy constraints can be obtained by varying the
action with respect to $\alpha_1$ and $\vartheta_1$. We find
\begin{align}
	2 \Mpl^2 H \alpha_1
	&=
		\dot \phi_I Q^I ,
	\label{eq:energy-constraint}\\
	-2 \Mpl^2 \frac{H}{a^2} \partial^2 \vartheta_1
	&=
		6 \Mpl^2 H^2 \alpha_1 -\alpha_1 \dot \phi_I \dot \phi^I 
		+ \dot \phi_I  \Dt Q^I
		+ V_{,I} Q^I .
	 \label{eq:momentum-constraint}
\end{align}
In writing 
Eqs.~\eqref{eq:2nd_order} and \eqref{eq:momentum-constraint}
we have used the spatial
Laplacian $\partial^2 \equiv \partial_i \partial_i$, with the convention
that spatial indices which are both written in the covariant position are
summed using the flat Euclidean metric $\delta^{ij}$.
Employing Eqs.~\eqref{eq:energy-constraint}
and~\eqref{eq:momentum-constraint} we can eliminate the metric
perturbations in $S_{(2)}$ to obtain
\begin{equation}
	\label{eq:S2}
	S_{(2)} =
	\frac{1}{2} \int \d^4 x \, a^3 \Big\{
		\Dt Q_I \Dt Q^I
		- h^{ij} \partial_i Q_I \partial_j Q^I
		- \M_{IJ} Q^I Q^J
	\Big\} ,
\end{equation}
where the symmetric mass matrix $\M_{IJ}$ satisfies
\begin{equation}
	\label{eq:mass}
	\M_{IJ} = V_{;IJ} - R_{LIJM} \dot \phi^L \dot \phi^M
	- \frac{1}{\Mpl^2 a^3} \Dt
	\left( \frac{a^3}{H} \dot \phi_I \dot \phi_J \right) .
\end{equation}
This is identical to the canonical case except for the covariant derivatives
and the term involving the Riemann tensor, which was first obtained by
Sasaki~\&~Stewart~\cite{Sasaki:1995aw}.
(See also Nakamura~\&~Stewart~\cite{Nakamura:1996da}
and Gong~\&~Stewart~\cite{Gong:2001he,Gong:2002cx}.)
We will find similar terms in the third-order action~\eqref{eq:s3sr} below.
Their meaning is not immediately clear because they imply that promotion to
curved field-space requires more than
`minimal coupling' to curvature.
In the mass matrix, the Riemann term changes the effective mass of modes
orthogonal to $\dot{\phi}$ but also alters the
coupling of these modes to each other.
We discuss these
effects in more detail 
in \S\ref{sec:evolution}.

\paragraph{Power spectrum at horizon-crossing.}
To compute the power spectrum and all higher $n$-point functions 
we must use the `in--in' formulation of quantum field theory, 
which entails doubling all field degrees of freedom.
For details, we refer to the
literature~\cite{Weinberg:2005vy,Seery:2007we,Chen:2010xka,Koyama:2010xj}.
In curved field-space an extra complication
is caused by the necessity to give each $n$-point
function the correct tensorial transformation properties.
As we describe in \S\ref{sec:third_order} below,
this is enforced by transport of the tangent-space basis
along the inflationary trajectory.

The power spectrum
at lowest order in $\M_{IJ}$ was calculated by
Sasaki~\&~Stewart~\cite{Sasaki:1995aw} and can be obtained
from~\eqref{eq:2pf-pp} or~\eqref{eq:2pf-mp}. Taking the equal-time limit
in either equation,
it follows that the
two-point function
evaluated
a little later than horizon-exit
satisfies
\begin{equation}
	\langle Q^I(\vect{k}_1) Q^J(\vect{k}_2) \rangle
	=
	(2\pi)^3 \delta(\vect{k}_1 + \vect{k}_2)
	\frac{H^2}{2k^3}
	\G^{IJ} .
	\label{eq:power-spectrum}
\end{equation}
This estimate becomes valid
when the decaying power-law terms
in~\eqref{eq:2pf-pp} and~\eqref{eq:2pf-mp} have become
negligible~\cite{Nalson:2011gc}.
However, because $Q^I$ is not infrared safe there are
growing terms at subleading order~\cite{Stewart:1993bc,Nakamura:1996da}.
Therefore~\eqref{eq:power-spectrum} soon becomes
untrustworthy,
and its validity extends only
for a very
narrow range of e-folds.
This is the problem of enhanced subleading terms
discussed in~\S\ref{sec:ir-safety}.
In Eq.~\eqref{eq:sep-univ-2pf} we use the separate universe
method to give a more accurate expression
which remains valid until late times.

\subsection{Third order} \label{sec:third_order}
The third order action is
\begin{equation}
\label{eq:s3-full}
\begin{split}
	S_{(3)} =
	\frac{1}{2} \int & \d^4 x \, a^3
	\bigg\{
		6 \Mpl^2 H^2 \alpha_1^3 
		+ 4 \Mpl^2 \frac{H}{a^2} \alpha_1^2 \partial^2 \vartheta_1 
		- \frac{\Mpl^2 \alpha_1}{a^4} \Big(
			\partial_i \partial_j \vartheta_1 \partial_i \partial_j \vartheta_1
			- \partial^2 \vartheta_1 \partial^2 \vartheta_1
		\Big)
		\\ & \mbox{}
		- \dot \phi_I \dot \phi^I \alpha_1^3
		+ 2 \alpha_1^2\dot \phi_I \Dt Q^I
		+ \frac{2}{a^2} \alpha_1 \dot{\phi}_I
			\partial_i \vartheta_1 \partial_i Q^I
		- \alpha_1 R^{L(IJ)M}\dot \phi_L \dot \phi_M Q_I Q_J
		\\ & \mbox{}
		- \alpha_1 \Big(
			\Dt Q_I \Dt Q^I
			+ \frac{1}{a^2} \partial_i Q_I \partial_i Q^I
		\Big)
		- \frac{2}{a^2} \partial_i \vartheta_1 \Dt Q_I \partial_i Q^I
		+ \frac{4}{3} R_{I(JK)L}\dot \phi^L \Dt Q^I Q^J Q^K 
		\\  & \mbox{}
		+ \frac{1}{3} R_{(I|LM|J;K)}\dot \phi^L \dot \phi^M Q^I Q^J Q^K
		- \frac{1}{3} V_{;(IJK)} Q^I Q^J Q^K
		- V_{;(IJ)} \alpha_1 Q^I Q^J
	\bigg\} .
\end{split}
\end{equation}
We have indicated the symmetric index combinations
picked out by each product
of the $Q^I$.%
	\footnote{Our symmetrization conventions are
	$2 A_{(IJ)} = A_{IJ} + A_{JI}$ and
	$6 A_{(IJK)} = A_{IJK} + \{ \text{5 perms} \}$.
	Bars delimit indices excluded from symmetrization.}
The lapse and shift can be eliminated 
using~\eqref{eq:energy-constraint}--\eqref{eq:momentum-constraint}.
The resulting expression is exact and does not
invoke an expansion in powers of slow-roll.
Nevertheless, the argument of~\S\ref{sec:ir-safety} implies that
we need only compute
lowest-order contributions
provided we evaluate the three-point function near horizon-crossing.
In this region
infrared divergences cannot enhance subleading terms.
Focusing only on the low-order contributions and
rewriting in terms of conformal time, we find~\cite{Gong:2011uw}
\begin{equation}
\begin{split}
	\label{eq:s3sr}
	S_{(3)} \supseteq \int \d^3 x \, \d \eta \, \bigg\{
		&
		- \frac{a^2}{4 \Mpl^2 H} \dot \phi^I Q_I \Deta Q^J \Deta Q_J 
		- \frac{a^2}{4 \Mpl^2 H} \dot \phi^I Q_I \partial_i Q^J \partial_i Q_J
		\\ & \mbox{} 
		+ \frac{a^2}{2 \Mpl^2 H} 
		\dot \phi^I \partial_i \partial^{-2} \Deta Q_I
		\partial_i Q_J \Deta Q^J 
		\\ & \mbox{} 
		+ \frac{2 a^3}{3} R_{I(JK)L}\dot \phi^L \Deta Q^I Q^J Q^K 
		+ \frac{a^4}{6} R_{(I|LM|J;K)}\dot \phi^L \dot \phi^M Q^I Q^J Q^K
	\bigg\} .
\end{split}
\end{equation}
The first two lines of~\eqref{eq:s3sr} are a covariantization of the action
obtained in flat field-space~\cite{Seery:2005gb}. 
We will sometimes describe them as the `canonical' terms.
Subleading corrections begin
at order $\Or(\dot{\phi}/H)^3$ and are negligible
unless enhanced by divergences.
The third line includes terms containing the Riemann tensor,
analogous to the Riemann term in the
mass matrix~\eqref{eq:mass}.
These were first obtained by Gong~\&~Tanaka~\cite{Gong:2011uw} 
and are a new feature associated with the curvature of field space. 
Unlike the lowest-order `canonical' terms they produce divergences. 
To track the influence of these as clearly as possible we have 
retained Riemann terms up to $\Or(\dot{\phi}/H)^2$.

Although operators involving the curvature at $\Or(\dot{\phi}/H)^2$ are
subleading, we can reliably compute contributions to the three-point
function at this level because Eq.~\eqref{eq:mass} shows that next-order
corrections from the propagator are themselves $\Or(\dot{\phi}/H)^2$, and
therefore enter the three-point function only at $\Or(\dot{\phi}/H)^3$.
The same is true for corrections from the scale factor and Hubble rate.
After expanding around the horizon-crossing time for a fiducial scale
$k_\ast$, the explicit $\Or(\dot{\phi}/H)^2$ term in~\eqref{eq:s3sr} 
is accompanied by one further contribution at the same order
from the time-dependence of $R_{I(JK)L}$.
[See Eqs.~\eqref{eq:riemann_3pf_expansion}
and~\eqref{eq:3pfv6}.]
We will retain both these terms when estimating the three-point function.
The advantage of doing so is that we can perform a more stringent test 
of our matching to the superhorizon evolution in~\S\ref{sec:evolution}.

\paragraph{Three-point function.}
In Appendix~\ref{sec:3pfcalcs} we calculate the contribution
to the three-point function from each of these operators. 
As explained in~\S\ref{sec:first_order}, our computation applies 
only to light field-space directions for which 
$\M_{IJ}$ can be neglected.

To express the result it will be necessary to
perform parallel transport along the inflationary
trajectory. This can be accomplished using the parallel propagator,%
	\footnote{In~\eqref{eq:covariant-expansion},
	to match the notation of Ref.~\cite{lrr-2011-7},
	we denoted
	the parallel propagator evaluated along a
	geodesic as ${\G_{I'}}^I$.
	We reserve the symbol ${\trajp}$ to indicate parallel propagation
	along an inflationary trajectory.}
\begin{equation}
	{\trajp^I}_{i}
	= \pathorder
		\exp \bigg(
			-\int_{\sigma}^{\tau} \d \eta
			\;
			{\Gamma^{I'}}_{M'N'} \frac{\d \phi^{M'}}{\d \eta}
		\bigg)
		{\G^{n}}_{i} ,
	\label{eq:trajectory-propagator}
\end{equation}
where the integral is computed along the inflationary trajectory between
conformal times $\sigma$ and $\tau$. The symbol `$\pathorder$'
denotes path ordering along this trajectory.
(More details and sample calculations
are given in Appendix~\ref{sec:3pfcalcs}.)
We refer to ${\trajp^I}_{i}$ as
the \emph{trajectory propagator}.
In writing~\eqref{eq:trajectory-propagator}
we have adopted a notation in which the index $I$ is associated with a basis
for the tangent space at
time $\tau$, whereas the index $i$ is associated with an
independent basis for the tangent space at $\sigma$.
Primed indices label the tangent space at a time
corresponding to the integration variable $\eta$.
The rightmost $N'$ index lies
at time $\sigma$ and contracts with ${\G^n}_i$.
Therefore ${\trajp^I}_i$ is a \emph{bitensor}:
it is an object with mixed indices~\cite{lrr-2011-7},
transforming like a contravariant vector at $\tau$
and a covariant vector at $\sigma$.
In what follows we set $\sigma$ to be the horizon-crossing time
for the reference scale $k_\ast$.

When computing $n$-point functions we typically measure time in
e-folds of expansion, evaluating
each $n$-point function at
$N = - \ln |k_\ast \tau|$
e-folds since the fiducial scale $k_*$ passed outside the horizon.
With these conventions we find
\begin{equation}
	\langle
		Q^I(\vect{k}_1)
		Q^J(\vect{k}_2)
		Q^K(\vect{k}_3)
	\rangle
	=
	(2\pi)^3 \delta(\vect{k}_1 + \vect{k}_2 + \vect{k}_3)
	\frac{H_\ast^4}{4 \prod_i k_i^3}
	{\trajp^I}_i {\trajp^J}_j {\trajp^K}_k
	A^{ijk}(N) .
	\label{eq:field_3pf}
\end{equation}
The bitensor $A^{ijk}(N)$ transforms as a scalar
at time $N$ and
a rank-three tensor 
at the time of horizon-crossing for $k_\ast$.
However, the three-point function
$\langle Q^I Q^J Q^K \rangle$ transforms as a rank-three tensor
at $N$. To match these transformation properties,
the propagators
${\trajp^I}_i$, ${\trajp^J}_j$ and
${\trajp^K}_k$
are necessary.
In Appendix~\ref{sec:3pfcalcs} we explain how they arise
in terms of Feynman diagrams.
We set $k_t = k_1 + k_2 + k_3$ to be the total scalar momentum and define
$\kappa^2 = \sum_{i < j} k_i k_j$.
Under the same conditions which were required for validity
of~\eqref{eq:power-spectrum}, we find
\begin{equation}
	\begin{split}
	A^{ijk}(N) = \mbox{} &
		\frac{1}{\Mpl^2} \frac{\dot{\phi}^i_\ast}{H_\ast} \G^{jk}_\ast
		\bigg(
			- 2 \frac{k_2^2 k_3^2}{k_t}
			+
			\frac{k_1}{2} \vect{k}_2 \cdot \vect{k}_3
		\bigg)
		\\ & \mbox{}
		+ \frac{4}{3} R^{i(jk)m}_\ast \frac{\dot{\phi}_m^\ast}{H_\ast}
		\left[
			k_1^3 \Big(
				\EulerGamma - N + \ln \frac{k_t}{k_\ast}
			\Big)
			- k_1^2 k_t
			+ \frac{k_1^2 k_2 k_3}{k_t}
		\right]
		\\ & \mbox{}
		+ \frac{1}{3} R^{(i|mn|j;k)}_\ast
		\frac{\dot{\phi}_m^\ast}{H_\ast}
		\frac{\dot{\phi}_n^\ast}{H_\ast}
		\left[
			k_1^3 \Big(
				N - \ln \frac{k_t}{k_\ast} - \EulerGamma - \frac{1}{3}
			\Big) 
			+ \frac{4}{9} k_t^3
			- k_t \kappa^2
		\right]
		\\ & \mbox{}
		- \frac{4}{3} R^{i(jk)m;n}_\ast
		\frac{\dot{\phi}_m^\ast}{H_\ast}
		\frac{\dot{\phi}_n^\ast}{H_\ast}
		\Bigg[
			\frac{k_1^3}{2} \Big(
				N^2
				- \EulerGamma^2 + \frac{\pi^2}{12}
				- \Big[
					2 \EulerGamma + \ln \frac{k_t}{k_\ast}
				\Big]
				\ln \frac{k_t}{k_\ast}
			\Big)
			\\ & \hspace{36mm} \mbox{}
			+ k_1^2 k_t \Big(
				\ln \frac{k_t}{k_\ast} + \EulerGamma - 1
			\Big)
			- \frac{k_1^2 k_2 k_3}{k_t}
			\Big(
				\EulerGamma + \ln \frac{k_t}{k_\ast}
			\Big)
		\Bigg]
		\\ & \mbox{}
		+ \text{cyclic} ,
	\end{split}
	\label{eq:bispectrum}
\end{equation}
where $\EulerGamma \approx 0.577$ is the Euler--Mascheroni constant.
The first line of~\eqref{eq:bispectrum}
covariantizes Eq.~(69) of Ref.~\cite{Seery:2005gb},
and subsequent lines arise from the Riemann terms in~\eqref{eq:s3sr}.
The sum over cyclic permutations includes the two permutations generated
by simultaneous exchange of 
$\{i$, $\vect{k}_1\}$ with either
$\{j$, $\vect{k}_2\}$ or $\{k$, $\vect{k}_3\}$.

\section{Evolution after horizon exit} \label{sec:evolution}

Terms in~\eqref{eq:bispectrum} involving $N$ are divergent in the late-time
limit $\tau \rightarrow 0$ and are responsible for spoiling infrared safety,
as described in~\S\ref{sec:ir-safety}.
They generate time evolution after horizon
exit~\cite{Zaldarriaga:2003my,Seery:2007wf,Seery:2010kh}
and rapidly invalidate the expressions derived
in~\S\ref{sec:n-pfs}.
In this section we
show that the divergent Riemann curvature terms
in~\eqref{eq:bispectrum}
have a geometrical origin and
explain how their contribution can be taken into account
using a covariant version of the `separate universe'
method~\cite{Lyth:2005fi,Seery:2012vj}.
The terms diverging like a single power of $N$ are
\begin{equation}
	A^{ijk}_{\text{1-$\log$}} =
		\frac{1}{3} N
		k_1^3
		\bigg(
			R^{(i|mn|j;k)}
			\frac{\dot{\phi}_{m}}{H}
			\frac{\dot{\phi}_{n}}{H}
			-
			4 {R^{i(jk)m}}
			\frac{\dot{\phi}_m}{H}
		\bigg)_*
		+
		\text{cyclic} .
	\label{eq:bispectrum-single-log}
\end{equation}
Eq.~\eqref{eq:bispectrum} also contains an explicit double logarithm 
(a term proportional to $N^2$)
\begin{equation}
	A^{ijk}_{\text{2-$\log$}} =
		-\frac{4}{6} N^2 k_1^3
		\bigg(		
		R^{i(jk)m;n}
		\frac{\dot{\phi}_m}{H}
		\frac{\dot{\phi}_n}{H}
		\bigg)_*
	+ \text{cyclic} .
	\label{eq:bispectrum-double-log}
\end{equation}
In addition, it implicitly contains terms of all powers in $N$
from higher-order corrections we have not evaluated.

\paragraph{Separate universe approach.}
Evolutionary effects on superhorizon scales
can be understood using causality
and classicality. After smoothing over
small-scale structure, widely separated regions locally evolve like an
unperturbed or `separate' Friedmann universe
\cite{Starobinsky:1982ee,Lyth:1984gv,Starobinsky:1986fxa}.
Lyth~\&~Rodr\'{\i}guez~\cite{Lyth:2005fi}
extended this method to 
$n$-point functions for $n \geq 3$. In their formulation, the principal
tool was a Taylor expansion of the background solutions in small deviations
from a chosen initial condition.

In Ref.~\cite{Seery:2012vj}, the separate universe approach was applied
without directly invoking this Taylor expansion. The inflationary
trajectories in phase space are interpreted
as the integral curves of a flow,
and can converge or disperse.
In Ref.~\cite{Seery:2012vj} it was shown that the Taylor coefficients
used by Lyth \& Rodr\'{\i}guez can be understood as (derivatives of) the
Jacobi fields which describe this dispersion.
The growth and decay of fluctuations, and the processes by which power is
transferred between modes, can be understood as the local dilation,
shear and twist of a narrow bundle of trajectories. 

\subsection{Jacobi equation for separate universes}
\label{sec:jacobi-separate}

The Jacobi method provides a simple way to implement the separate
universe approach in curved field-space.
Consider two separate
universes, with slightly displaced initial conditions, which correspond to
neighbouring trajectories on phase space.
The displacement between these trajectories can be described covariantly using
a tangent-space vector $Q^I$ (the `connecting vector') as described
in~\S\ref{sec:covariant-perts}.

\paragraph{Deviation equation.}
Each universe evolves according to the field equation
\begin{equation}
	\frac{1}{3} \DN^2 \vp^I +
	\left( 1 - \frac{\epsilon}{3} \right) \DN \vp^I = u^I ,
	\label{eq:trajectory-eq}
\end{equation}
where $u_I = - V_{,I} / 3H^2$. 
Under the slow-roll approximation the acceleration term $\DN^2 \vp^I$
is negligible along each trajectory.
In flat field-space this means that the change in acceleration term between
neighbouring trajectories also contributes at higher-order in slow-roll.
On curved field-space this is no longer true because
derivatives do not
commute. Therefore we must retain the acceleration term when studying how
trajectories disperse.

The evolution of $Q^I$ can be determined by making a Taylor expansion
of~\eqref{eq:trajectory-eq} along a geodesic connecting the adjacent
trajectories, as in \S\ref{sec:covariant-perts}.
To describe evolution of the two- and three-point functions we require this
expansion up to second-order. Dropping the explicit $\Or(\epsilon)$ term,
which can contribute only at higher order in the slow-roll expansion,
and discarding a common factor of the parallel propagator we find
\begin{equation}
	\left( \Dl + \frac{1}{2} \Dl^2 \right)
	\left(
		\frac{1}{3}	\DN^2 \vp^I
		+
		\DN \vp^I
	\right)
	=
	{u^I}_{;J} Q^J
	+ \frac{1}{2} {u^I}_{;JK} Q^J Q^K .
	\label{eq:taylor-jacobi}
\end{equation}
Using the Ricci identity to commute $\Dl$ with $\DN$
and employing the Bianchi identities to symmetrize resulting
curvature terms, we conclude that $Q^I$ evolves according to the
Jacobi or `deviation' equation
\begin{equation}
	\label{eq:jacobi-eq}
	\DN Q^I
	= {w^I}_J Q^J + \frac{1}{2} {w^I}_{(JK)} Q^J Q^K + \cdots ,
\end{equation}
where the coefficients ${w^I}_J$ and ${w^I}_{(JK)}$ satisfy
\begin{eqnarray}
	\label{eq:jacobi-w2}
	w_{IJ} & = & u_{(I;J)} 
	+ \frac{1}{3} R_{L(IJ)M}
	\frac{\dot{\phi}^L}{H}
	\frac{\dot{\phi}^M}{H} , \\
	\label{eq:jacobi-w3}
	w_{I(JK)} & = & u_{(I;JK)}
		+ \frac{1}{3} \bigg(
			R_{(I|LM|J;K)} \frac{\dot{\phi}^L}{H} \frac{\dot{\phi}^M}{H}
			- 4 R_{I(JK)L} \frac{\dot{\phi}^L}{H}
		\bigg) .
\end{eqnarray}
To obtain~\eqref{eq:jacobi-w2}--\eqref{eq:jacobi-w3} we have
imposed the slow-roll approximation to determine $\DN^2 Q^I$ in terms
of $\DN Q^I$.
As usual, the background trajectory is denoted $\phi^I(t)$.
All curvature quantities
and derivatives of $u^I$ are evaluated on this trajectory and therefore
powers of slow-roll can be counted in the usual way.
Because we have used
the slow-roll approximation,
Eqs.~\eqref{eq:jacobi-w2}--\eqref{eq:jacobi-w3}
are trustworthy only to lowest slow-roll order
in the derivatives of $u^I$,
and to $\Or(\dot{\phi}/H)^2$ in terms multiplying the Riemann tensor
and its derivatives.
This accuracy is sufficient to make
a comparison with the divergent terms retained in~\S\ref{sec:third_order}.

Although both terms in~\eqref{eq:jacobi-w2} are automatically
symmetric under exchange of $IJ$ we have indicated this explicitly.
However, $w_{IJK}$ is symmetric only on $JK$.
This is different to the case of flat field-space, where terms involving the
Riemann tensor are absent and each coefficient on the right-hand side of the
Jacobi equation is always a symmetric combination of partial derivatives.
When writing ${w^I}_{(JK)}$ we add redundant
brackets to emphasize 
this symmetry.

\paragraph{Time-evolution operators.}
The Jacobi equation~\eqref{eq:jacobi-eq} is a first-order differential equation, and therefore its solution
can be expanded in powers of the initial conditions
$Q^m_*$,%
	\footnote{The quantities ${\bigamma^I}_m$ and
	${\bigamma^I}_{mn}$ were written
	${\Gamma^I}_m$ and ${\Gamma^I}_{mn}$ in
	Refs.~\cite{Seery:2012vj,Anderson:2012em}.
	In this paper we reserve $\Gamma$ to mean the Levi-Civita
	connection compatible with the field-space metric
	$\G^{IJ}$.}
\begin{equation}
	Q^I = {\bigamma^I}_m Q^m_*
		+ \frac{1}{2} {\bigamma^I}_{(mn)} Q^m_* Q^n_* + \cdots .
	\label{eq:jacobi-soln}
\end{equation}
To write~\eqref{eq:jacobi-soln} we have used the index
convention introduced in~\S\ref{sec:third_order}.
The fluctuation $Q^I$ is evaluated at some late time $N$,
and its index $I$ transforms as a contravariant vector in the
tangent space at this time.
Conversely, $Q^m_*$ and its index $m$
transform as a contravariant vector at an earlier time $N_*$.
Like the trajectory propagator~\eqref{eq:trajectory-propagator},
the coefficients ${\bigamma^I}_m$ and
${\bigamma^I}_{(mn)}$ are bitensors.
In particular, ${\bigamma^I}_m$ transforms like a
contravariant vector on $I$ and a covariant vector on $m$,
whereas ${\bigamma^I}_{(mn)}$
transforms like a contravariant vector in the
tangent space at $N$ and a covariant rank-two tensor in the tangent space
at $N_\ast$.
The initial conditions require ${\bigamma^I}_m = \delta^I_m$ and 
${\bigamma^I}_{(mn)} = 0$ when $N = N_\ast$.

Eq.~\eqref{eq:jacobi-soln} is a solution to the Jacobi
equation~\eqref{eq:jacobi-eq} provided the $\bigamma$ coefficients satisfy
\begin{align}
	\DN {\bigamma^I}_m & =
		{w^I}_J {\bigamma^J}_m ,
	\label{eq:gamma2-eq}
	\\
	\DN {\bigamma^I}_{(mn)}
	& = {w^I}_J {\bigamma^J}_{(mn)}
	+ {w^I}_{(JK)} {\bigamma^J}_m {\bigamma^K}_n .
	\label{eq:gamma3-eq}
\end{align}
(Recall that $N$ in the derivative $\DN$ is not
a field-space index, but the number of e-folds.)
Eq.~\eqref{eq:jacobi-soln}
can be summarized as the Taylor expansion
of $Q^I$ in terms of its value at some earlier time $N_\ast$,
and defines
a `separate universe' approximation for
curved field-space.
We describe the
coefficients ${\bigamma^I}_{(m \cdots n)}$
collectively as
`time-evolution operators'.
They are covariant analogues of the coefficients
$\partial \phi^I / \partial \phi^m_\ast$
(and its higher derivatives) which occur when
applying the separate-universe method in flat
field-space~\cite{DeWittMorette:1976up,Lewandowski:1993zq}.
We could obtain them
by solving $\vp^I$ and
using~\eqref{eq:covariant-expansion} to compute its derivatives with
respect to the initial conditions,%
	\footnote{To reproduce all information
	in~\eqref{eq:jacobi-w2}--\eqref{eq:jacobi-w3}
	it would be necessary to retain the connection term from
	$\Dt^2 \vp^I$.}
but in practice it is much easier to
integrate~\eqref{eq:gamma2-eq}--\eqref{eq:gamma3-eq} directly.

\paragraph{Divergences.}
We now show that~\eqref{eq:jacobi-eq}--\eqref{eq:jacobi-w3} reproduce the
divergences of~\eqref{eq:bispectrum}. The argument
is similar to that of
Zaldarriaga~\cite{Zaldarriaga:2003my}. 
Solving Eqs.~\eqref{eq:gamma2-eq}--\eqref{eq:gamma3-eq} perturbatively
yields a power series in $N$. The lowest-order terms are
\begin{align}
	\label{eq:gamma2-soln}
	{\bigamma^I}_m & =
		{\trajp^I}_m
		+ {\trajp^I}_i \Big[{w^i}_m \Big]_\ast N
		+ \frac{1}{2} {\trajp^I}_i
			\Big[{w^i}_k {w^k}_m + \DN {w^i}_m \Big]_\ast N^2
		+ \cdots , \\
		\label{eq:gamma3-soln}
	{\bigamma^I}_{(mn)} & =
		{\trajp^I}_i \Big[ {w^i}_{(mn)} \Big]_* N
		\\ & \qquad \nonumber 
		+ \frac{1}{2} {\trajp^I}_i \Big[
			\DN {w^i}_{(mn)}
			+ {w^i}_k {w^k}_{(mn)}		
			+ {w^i}_{(mk)} {w^k}_n
			+ {w^i}_{(nk)} {w^k}_m
		\Big]_\ast N^2
		+ \cdots ,
\end{align}
where $N = -\ln|k_\ast \tau|$ as above.

Eq.~\eqref{eq:gamma2-soln}
shows that
the time-evolution operator
${\bigamma^I}_m$ can be understood as
a modification of the trajectory propagator to include the effect of
time-dependence along the inflationary trajectory in addition to
parallel transport.
This follows because
the trajectory propagator ${\trajp^I}_m$
satisfies~\eqref{eq:gamma2-eq} with ${w^I}_J = 0$.

At linear order in $N$,
the two- and three-point functions following
from~\eqref{eq:gamma2-soln}--\eqref{eq:gamma3-soln} are
\begin{align}
	\label{eq:divergent-2pf}
	\langle Q^I(\vect{k}_1) Q^J(\vect{k}_2) \rangle
	& \supseteq
	(2\pi)^3 \delta(\vect{k}_1 + \vect{k}_2)
	\frac{N H_*^2}{k^3}
	{\trajp^I}_{i}
	{\trajp^J}_{j}
	{w}^{ij}_\ast , \\
	\label{eq:divergent-3pf}
	\langle Q^I(\vect{k}_1) Q^J(\vect{k}_2) Q^K(\vect{k}_3) \rangle
	& \supseteq
	(2\pi)^3 \delta(\vect{k}_1 + \vect{k}_2 + \vect{k}_3)
	\frac{N H_*^4}{4\prod_i k_i^3} 
		{\trajp^I}_i {\trajp^J}_j {\trajp^K}_k
		w^{i(jk)}_\ast k_1^3
		+ \text{cyclic} .
\end{align}
The symmetry properties of $w^{ij}$ and $w^{i(jk)}$ ensure that these
expressions are symmetric under simultaneous permutations of the indices
$I$, $J$, $K$ and their associated momenta $\vect{k}_1$, $\vect{k}_2$,
$\vect{k}_3$.
At quadratic order in $N$ and lowest order in slow-roll
there is a contribution only from the $\DN {w^i}_{(mn)}$ term
in~\eqref{eq:gamma3-soln}.
This gives
\begin{equation}
\label{eq:double-divergent-3pf}
\begin{split}
	\langle Q^I(\vect{k}_1) Q^J(\vect{k}_2) Q^K(\vect{k}_3) \rangle
	\supseteq \mbox{}
	&
	(2\pi)^3 \delta(\vect{k}_1 + \vect{k}_2 + \vect{k}_3)
	\frac{N^2 H_\ast^4}{4\prod_i k_i^3}
	\\ & \mbox{} \times
		{\trajp^I}_i {\trajp^J}_j {\trajp^K}_k
		\bigg(
			-\frac{4}{6} R^{i(jk)m;n}
			\frac{\dot{\phi}_m}{H}
			\frac{\dot{\phi}_n}{H}
		\bigg)_* k_1^3
		+ \text{cyclic} .
\end{split}
\end{equation}

It can be checked that~\eqref{eq:divergent-2pf} reproduces the divergence
in the two-point function (including the term involving the Riemann tensor)
found by Nakamura \& Stewart~\cite{Nakamura:1996da}.%
	\footnote{In Refs.~\cite{Nakamura:1996da,Gong:2001he,Gong:2002cx}
	the factors of ${\trajp^I}_i$ were omitted.}
Comparing~\eqref{eq:divergent-3pf} and~\eqref{eq:jacobi-w3}, it can also
be checked that the terms in $w^{i(jk)}$ involving the Riemann tensor
reproduce the divergence in~\eqref{eq:bispectrum-single-log}.
It was to enable a nontrivial check of this matching that we elected to
keep divergences up to $\Or(\dot{\phi}/H)^2$ in the Riemann-tensor terms
of~\eqref{eq:s3sr}.
Finally, comparing~\eqref{eq:double-divergent-3pf}
and~\eqref{eq:bispectrum-double-log}
it can be checked that the lowest-order
double-logarithmic divergence
is also correctly reproduced.
At the accuracy of our present calculation it is not possible to check
whether the divergences proportional to $u_{(i;jk)}$ also agree.
These are a covariantized version of the divergences which appear in
flat field-space.
Since the same is true for the `canonical' operators in the
Lagrangian~\eqref{eq:s3-full} we should expect agreement.

In principle, higher-order terms
in $N$ and $\dot{\phi}/H$ could be
retained in the perturbative
expansions~\eqref{eq:gamma2-soln}--\eqref{eq:gamma3-soln}, which would enable
a check of matching at all orders.
Although such a check would be interesting, the matching at
single- and double-logarithm
order provides no reason to believe it would fail. We will not attempt it in
this paper.

\subsection{Transport equations}
\label{sec:transport-equations}
The 
time-evolution operators
enable us to determine 
each $n$-point function after horizon exit.
Translating the formulae of Lyth~\&~Rodr\'{\i}guez, we obtain
\begin{equation}
	\langle
		Q^I(\vect{k}_1)
		Q^J(\vect{k}_2)
	\rangle
	=
		{\bigamma^I}_m
		{\bigamma^J}_n
		\langle
			Q^m(\vect{k}_1)
			Q^n(\vect{k}_2)
		\rangle_*
	\label{eq:sep-univ-2pf}
\end{equation}
and
\begin{equation}
\begin{split}
	\langle
		Q^I(\vect{k}_1)
		&
		Q^J(\vect{k}_2)
		Q^K(\vect{k}_3)
	\rangle
	=
		{\bigamma^I}_l
		{\bigamma^J}_m
		{\bigamma^K}_n
		\langle
			Q^l(\vect{k}_1)
			Q^m(\vect{k}_2)
			Q^n(\vect{k}_3)
		\rangle_*
	\\
	& \mbox{}
		+
		{\bigamma^I}_{(lm)}
		{\bigamma^K}_{r}
		{\bigamma^J}_{s}
		\int \frac{\d^3 q}{(2\pi)^3}
		\langle
			Q^l(\vect{k}_1 - \vect{q})
			Q^r(\vect{k}_2)
		\rangle_*
		\langle
			Q^m(\vect{q})
			Q^s(\vect{k}_3)
		\rangle_*
		+
		\text{cyclic} ,
\end{split}
\label{eq:sep-univ-3pf}
\end{equation}
where `cyclic' denotes the two permutations of the second line
in~\eqref{eq:sep-univ-3pf} obtained by exchanging $\{ I, \vect{k}_1 \}$ with
$\{ J, \vect{k}_2 \}$ or $\{ K, \vect{k}_3 \}$.
When there is no time evolution and
${w^I}_J = 0$,
Eq.~\eqref{eq:sep-univ-3pf} reproduces~\eqref{eq:field_3pf}.
In combination, Eqs.~\eqref{eq:bispectrum},
\eqref{eq:jacobi-w2}--\eqref{eq:jacobi-w3},
\eqref{eq:gamma2-eq}--\eqref{eq:gamma3-eq}
and~\eqref{eq:sep-univ-2pf}--\eqref{eq:sep-univ-3pf}
constitute the principal results of this paper.
For comparison with observation it 
only remains
to make a gauge transformation from $Q^I$ to the curvature perturbation
$\zeta$, for which we supply the relevant formulae in \S\ref{sec:observables}.

Up to this point we have worked in a holonomic frame derived from the
field-space coordinates, but other possibilities exist.
Since an $n$-point function of the $Q^m$ evaluated at $N_\ast$ transforms as
a rank-$n$ tensor in the tangent-space at time $N_\ast$, 
Eqs.~\eqref{eq:sep-univ-2pf}--\eqref{eq:sep-univ-3pf} are manifestly covariant.
As a result, we are free to select a basis for the tangent space
independently at the early and late times $N_\ast$ and $N$. 

\paragraph{Equations for two- and three-point functions.}
The approach given above is simple and emphasizes its similarity with
familiar $\delta N$ methods, but it is also possible to write evolution
equations for the $n$-point functions directly. 
In Ref.~\cite{Seery:2012vj} these were described as \emph{transport equations}.

We write the two-point function as
\begin{equation}
	\langle
		Q^I(\vect{k}_1)
		Q^J(\vect{k}_2)
	\rangle
	=
	(2\pi)^3
	\delta(\vect{k}_1 + \vect{k}_2)
	\frac{\Sigma^{IJ}}{2k^3} ,
\end{equation}
where $\Sigma^{IJ}$ is symmetric.
The amplitude of the local mode of the three-point
function can be parametrized
\begin{equation}
	\langle
		Q^I(\vect{k}_1)
		Q^J(\vect{k}_2)
		Q^K(\vect{k}_3)
	\rangle
	=
	(2\pi)^3
	\delta(\vect{k}_1 + \vect{k}_2 + \vect{k}_3)
	\left[
		\frac{\alpha^{I(JK)}}{k_2^3 k_3^3}
		+
		\frac{\alpha^{J(IK)}}{k_1^3 k_3^3}
		+
		\frac{\alpha^{K(IJ)}}{k_1^3 k_2^3}
	\right] .
\end{equation}
Direct differentiation followed by use of~\eqref{eq:jacobi-eq}
yields the equations
\begin{align}
	\DN \Sigma^{IJ}
	& =
	{w^I}_L \Sigma^{LJ} + {w^J}_L \Sigma^{LI} + \cdots ,
	\label{eq:2pf-transport}
	\\
	\DN \alpha^{I(JK)}
	& =
	{w^I}_L \alpha^{L(JK)}
	+ {w^J}_L \alpha^{I(LK)}
	+ {w^K}_L \alpha^{I(JL)}
	+ {w^I}_{(LM)} \Sigma^{LJ} \Sigma^{MK}
	+ \cdots ,
	\label{eq:3pf-transport}
\end{align}
where the omitted terms involve higher-order correlation functions and are
negligible in typical inflationary theories.
Following the method described in Ref.~\cite{Seery:2012vj} it can be
verified that~\eqref{eq:2pf-transport}--\eqref{eq:3pf-transport} reproduce
Eqs.~\eqref{eq:sep-univ-2pf}--\eqref{eq:sep-univ-3pf}.

\subsubsection{Interpretation of Riemann terms}
\label{sec:riemann-terms}

It is now possible to understand the significance of
the interactions in~\eqref{eq:bispectrum}
mediated by the Riemann curvature, and 
the infrared divergences to which they give rise.
They correspond directly to those terms in the separate-universe
Jacobi equation~\eqref{eq:jacobi-eq} which measure deviation
between nearby trajectories. The derivation
of~\S\ref{sec:jacobi-separate} makes clear that this effect
is entirely analogous to geodesic deviation between freely-falling
observers in curved spacetime.

These new sources of time dependence arise
mathematically from
tidal effects in field space.
Their physical meaning can be understood as follows. An
initial perturbation $Q^m_*$ generically represents a mix of
adiabatic and isocurvature fluctuations.
The isocurvature fluctuations
differentiate between `separate universes',
and correspond to a choice of inflationary trajectory
measured from the fiducial trajectory at $Q^I = 0$.
As these trajectories flow over field-space the proper distance between
them will vary depending on the metric $\G_{IJ}$.
Each trajectory evolves independently
in the absence of gradient terms which couple spatially separated regions.
Therefore the isocurvature component of $Q^I$ represents a
displacement to a fixed trajectory and
must respond to this variation.

These effects 
generate new sources of
dilation and shear which contribute to the redistribution of power between
adjacent trajectories. In addition, the \emph{nonlinear}
terms in ${w^I}_{(JK)}$
generate new sources of non-Gaussianity.
The precise form of these nonlinear terms is unfamiliar only because
they are irrelevant in
familiar applications of geodesic deviation---such as the focusing and
defocusing of a bundle of light rays, where the connecting vector can be taken to be infinitesimal.

The Riemann tensor is antisymmetric on its first and second pairs of
indices. Since $\dot{\phi}^I$ is proportional to the tangent to the curve,
we conclude that the Riemann contribution
to ${w^I}_J$ is zero when either index is
aligned with the adiabatic direction.
Gong~\&~Tanaka emphasized that this leads
only to new couplings between isocurvature modes~\cite{Gong:2011uw}.
Eq.~\eqref{eq:2pf-transport} shows that
these couplings
influence how the
isocurvature modes share power between themselves, but do not cause power
to flow between the isocurvature and adiabatic directions. 
Such a flow must be mediated by the potential through $u_{(I;J)}$.

In the special case where the trajectory is an exact geodesic
its tangent vector is parallel-transported proportional to itself.
In this case, the adiabatic
mode decouples completely and no power flows to or from it.

\subsection{Gauge transformation to the curvature perturbation}
\label{sec:observables}

For comparison with microwave background experiments
or galaxy surveys we must compute the $n$-point functions of the primordial
curvature perturbation, $\zeta$.
This is a measure of the local excess expansion between
a spatially flat hypersurface and a uniform density hypersurface with which it
coincides on average. In curved field-space
the computation can be performed economically using
the method of Ref.~\cite{Anderson:2012em}.
We expand $N$ as a function of the density $\rho$.
Taking $\Delta \rho$ to be the displacement from a point of fixed density
$\rho_c$ to an arbitrary initial location, we find
\begin{equation}
	\Delta N =
		\frac{\d N}{\d \rho}
		\Delta \rho
		+
		\frac{1}{2}
		\frac{\d^2 N}{\d \rho^2}
		(\Delta \rho)^2
		+
		\cdots
		.
	\label{eq:primitive-deltaN}
\end{equation}
To determine the variation of~\eqref{eq:primitive-deltaN}
under a change in the initial location we expand along a
geodesic, as in~\S\ref{sec:covariant-perts},
along which both $\Delta\rho$ and the differential coefficients will vary.
The variation of $\Delta\rho$ satisfies
\begin{equation}
	\delta(\Delta \rho) =
		- V_{;I} Q^I - \frac{1}{2} V_{;IJ} Q^I Q^J + \cdots .
\end{equation}
Therefore, up to second order, we can express $\zeta$ as
\begin{equation}
	\zeta
		= \delta(\Delta N)
		= N_I Q^I + \frac{1}{2} N_{IJ} Q^I Q^J + \cdots .
	\label{eq:zeta-def}
\end{equation}
The coefficients $N_I$ and $N_{IJ}$ satisfy
\begin{align}
	\label{eq:N1-gauge}
	N_I & = - \frac{\d N}{\d \rho} V_{;I} , \\
	\label{eq:N2-gauge}
	N_{IJ} & = - \frac{\d N}{\d \rho} V_{;IJ}
		+ \frac{\d^2 N}{\d\rho^2} V_{;I} V_{;J}
		+ \frac{1}{\Mpl^2} \big(
			A_{I;J} + A_{J;I}
		\big) ,
\end{align}
where 
\begin{align}
	A_I & =
		\frac{V_{;I}}{V^{;J}V_{;J}}
		- \frac{2V}{(V^{;J}V_{;J})^2} V^{;K}V_{;IK} , \\
	\frac{\d N}{\d \rho} &=
		- \frac{1}{\Mpl^2} \frac{V}{V^{;I}V_{;I}} , \\
	\frac{\d^2 N}{\d \rho^2} &=
		- \frac{1}{\Mpl^2} \frac{1}{V^{;I}V_{;I}}
		+ \frac{2}{\Mpl^2} \frac{V}{(V^{;I}V_{;I})^3} V^{;J}V^{;K}V_{;JK} .
\end{align}

\paragraph{$\delta N$ coefficients.}
Eqs.~\eqref{eq:N1-gauge}--\eqref{eq:N2-gauge}
are defined at a single point
in field space;
they are not bilocal in the sense of the coefficients ${\bigamma^I}_m$
and ${\bigamma^I}_{(mn)}$. 
We can obtain analogues of these bilocal coefficients
using~\eqref{eq:jacobi-soln} to relate the $Q^I$
to their values
at horizon-crossing.
This yields
\begin{equation}
	\zeta(N)
	=
		\biN_m Q^m_\ast
		+
		\frac{1}{2} \biN_{(mn)} Q^m_\ast Q^n_\ast
		+
		\cdots ,
	\label{eq:covariant-deltaN}
\end{equation}
where $\biN_m$ and $\biN_{(mn)}$
transform as scalars in the tangent space at
$N$, and (respectively) rank-one and rank-two tensors in the tangent space
at $N_\ast$.
We describe them as `$\delta N$ coefficients'.
They satisfy
\begin{align}
	\label{eq:N1}
	\biN_m & = N_I {\bigamma^I}_m , \\
	\label{eq:N2}
	\biN_{(mn)} & = N_I {\bigamma^I}_{(mn)} +
		N_I N_J {\bigamma^I}_m {\bigamma^J}_n .
\end{align}
It follows that the two- and three-point functions
of $\zeta$ are determined by
\begin{equation}
	\langle
		\zeta(\vect{k}_1)
		\zeta(\vect{k}_2)
	\rangle
	=
		\biN_m \biN_n
		\langle
			Q^m(\vect{k}_1)
			Q^n(\vect{k}_2)
		\rangle_* ,
	\label{eq:zeta-2pf}
\end{equation}
and
\begin{equation}
\begin{split}
	\langle
		\zeta(\vect{k}_1)
		&
		\zeta(\vect{k}_2)
		\zeta(\vect{k}_3)
	\rangle
	=
		\biN_l \biN_m \biN_n
		\langle
			Q^l(\vect{k}_1)
			Q^m(\vect{k}_2)
			Q^n(\vect{k}_3)
		\rangle_* 
	\\
	& \mbox{} +
		\biN_{(lm)} \biN_r \biN_s
		\int \frac{\d^3 q}{(2\pi)^3}
		\langle
			Q^l(\vect{k}_1 - \vect{q})
			Q^r(\vect{k}_2)
		\rangle_*
		\langle
			Q^m(\vect{q})
			Q^s(\vect{k}_3)
		\rangle_*
		+ \text{cyclic} ,
	\label{eq:zeta-3pf}
\end{split}
\end{equation}
where `cyclic' indicates the usual combination of permutations,
as in Eq.~\eqref{eq:sep-univ-3pf}.

Eq.~\eqref{eq:N2} implies that
$\biN_{(mn)}$ is the covariant derivative
of $\biN_{m}$ with respect to a change in the initial conditions.
Therefore~\eqref{eq:covariant-deltaN} agrees with the
covariant $\delta N$ expansion discussed by Saffin~\cite{Saffin:2012et}.
A similar expansion has already been used
by Peterson~\&~Tegmark~\cite{Peterson:2011yt}.

\paragraph{Observable quantities.}
The statistical properties of $\zeta$ are typically expressed in
terms of its spectrum and bispectrum
\begin{align}
	\label{eq:Pz}
	\langle
		\zeta_{\vect{k}_1}
		\zeta_{\vect{k}_2}
	\rangle
	& \equiv
	(2\pi)^3 \delta(\vect{k}_1 + \vect{k}_2) P_\zeta(k_1) ,
	\\
	\label{eq:Bz}
	\langle
		\zeta_{\vect{k}_1}
		\zeta_{\vect{k}_2}
		\zeta_{\vect{k}_3}
	\rangle
	& \equiv
	(2\pi)^3 \delta(\vect{k}_1 + \vect{k}_2 + \vect{k}_3)
	B_\zeta(k_1, k_2, k_3) .
\end{align}
Constraints on the
spectrum are
expressed in terms
of a dimensionless
quantity
$\P_\zeta$
\begin{equation}
	\P_\zeta(k) = \frac{k^3}{2\pi^2} P_\zeta(k)
	= \biN_m \biN_n \G^{mn}(k) \frac{H(k)^2}{4\pi^2} ,
	\label{eq:p-zeta}
\end{equation}
where the argument $k$ denotes evaluation at the horizon-crossing time for $k$.

The bispectrum contains considerably more information.
Its terms can be divided into two types:
first, those which come from interference effects
involving decaying modes
near horizon exit;
and second, those arising from interactions between only growing modes
far outside the horizon.
The first type can have arbitrary dependence on the $k$-modes
$k_1$, $k_2$, $k_3$ which appear in the three-point correlation
function.
The second type appear only in the `local' combination
$(k_1^3 k_2^3)^{-1}$ or its permutations.
With canonical kinetic terms,
applying global slow-roll conditions to the potential
and assuming that only light fields contribute to $\zeta$,
Lyth~\&~Zaballa~\cite{Lyth:2005qj}
showed that the result of Ref.~\cite{Seery:2005gb}
implies only the `local' bispectrum shape can be observable.

With a nontrivial field-space metric, interactions involving the
Riemann tensor may change this conclusion.
Inspection of~\eqref{eq:bispectrum} shows that these interactions
contribute terms of both types.
Depending on the field-space curvature it is possible that
`nonlocal' contributions in~\eqref{eq:bispectrum} could be enhanced,
but to determine whether this happens would require an extension of the
analysis in Refs.~\cite{Lyth:2005qj,Vernizzi:2006ve}.
We leave this interesting question for future work.
On the other hand these terms certainly modify the evolution of the
amplitude of each local shape, as discussed
in~\S\S\ref{sec:jacobi-separate}--\ref{sec:transport-equations}.

Where the Riemann curvature is sufficiently
small to make nonlocal contributions negligible,
Eq.~\eqref{eq:zeta-3pf} yields an analogue of the familiar
`$\delta N$' formula for the amplitude of the local bispectrum,
\begin{equation}
	\fnl^{\text{local}}
	\approx
		\frac{5}{6} \frac{\biN_m \biN_n \biN^{(mn)}}{(\biN_r \biN^r)^2} ,
	\label{eq:fNL}
\end{equation}
where $\biN_m$ and $\biN_{(mn)}$ were defined in~\eqref{eq:N1}--\eqref{eq:N2}.
In this case, Eq.~\eqref{eq:bispectrum} can be interpreted to mean
(as in the case of canonical kinetic terms) that
the bispectrum generated at horizon exit is negligible:
subsequent time evolution is necessary
to generate an observable non-Gaussian signal.

\subsubsection{Backwards formalism}
To track the evolution of mixed two- and three-point functions for
the complete set of fluctuations, including isocurvature
modes, it is necessary to solve for
all components of ${\bigamma^I}_{m}$ and ${\bigamma^I}_{(mn)}$.
In an $\N$-field model, the linear coefficient
${\bigamma^I}_m$ has $\N^2$ independent components and
the quadratic coefficient ${\bigamma^I}_{(mn)}$
has $\N^2(\N+1)/2$ independent components.
But for some purposes we may require only
the two- and three-point functions for $\zeta$
given by~\eqref{eq:zeta-2pf}--\eqref{eq:zeta-3pf}.
If so, we can reduce the computational complexity by
tracking only the $\N$ components of $\biN_n$
and the $\N(\N+1)/2$ components of $\biN_{(mn)}$.

An evolution equation for $\biN_n$ was given by
Yokoyama, Suyama \& Tanaka~\cite{Yokoyama:2007uu,Yokoyama:2007dw}.
Assuming the local shape dominates, only the combination
$\biN_m \biN_{(mn)} \biN_n$
is required. Yokoyama, Suyama \& Tanaka
supplied an integral expression from which this could be computed.
In Ref.~\cite{Seery:2012vj} this was extended to
an explicit evolution equation for $\biN_{(mn)}$.
Equivalent expressions were later given by
Mazumdar \& Wang~\cite{Mazumdar:2012jj}.

Eq.~\eqref{eq:gamma2-eq} expresses the evolution
of ${\bigamma^I}_m$ with $N$.
It can be verified that its
evolution with $N_\ast$ satisfies an analogous equation
\begin{equation}
	\mathcal{D}_{N_*}{\bigamma^I}_m
	=
	-{\bigamma^I}_n {w^n}_m .
	\label{eq:gamma2-backwards}
\end{equation}
Here, the covariant derivative applies to tangent-space indices
at $N_\ast$ and therefore operates only on $m$. The
index $I$ labels a tangent-space basis at $N$ and is inert.
Using~\eqref{eq:gamma2-backwards}
together with~\eqref{eq:N1}--\eqref{eq:N2} we conclude
\begin{align}
	\label{eq:N1-backwards}
	\mathcal{D}_{N_*} \biN_n
	& =
		- \biN_m {w^m}_n , \\
	\mathcal{D}_{N_*} \biN_{(mn)}
	\label{eq:N2-backwards}
	& =
		- \biN_{(rn)} {w^r}_m
		- \biN_{(mr)} {w^r}_n
		- \biN_{r} {w^r}_{(mn)} .
\end{align}
Like Eqs.~\eqref{eq:gamma2-eq}--\eqref{eq:gamma3-eq} these
are covariantized versions of the evolution equations in flat field-space,
using the correct coefficients ${w^m}_n$ and ${w^m}_{(rs)}$
which appear in the Jacobi equation.

Eqs.~\eqref{eq:N1-backwards}--\eqref{eq:N2-backwards}
should be solved by fixing $N$ to be the late time of interest.
During inflation the initial conditions would correspond to
\eqref{eq:N1-gauge}--\eqref{eq:N2-gauge}.
One should then integrate backwards until $N_\ast$ corresponds
to the horizon-crossing time for the fiducial scale $k_\ast$.

\section{Non-minimally coupled models} \label{sec:action}

In a multiple-field model with potential $V$
and non-minimal coupling to the Ricci scalar,
the action in the Jordan frame can be written as
\begin{equation}
	S = \int \d^4 x \, \sqrt{-\hat{g}} \left[
		\frac{1}{2} \Mpl^2 \, f(\vp) \hat{R}
		+ \hat{X}
		- \hat{V}(\vp)
	\right] ,
\end{equation}
where $f(\vp)$
is positive definite but otherwise arbitrary.
We distinguish Jordan frame quantities with a circumflex.
The kinetic energy is $\hat{X}$,
assumed to be an arbitrary second-order combination of the field derivatives
in the form
\begin{equation}
	\hat{X} = - \frac{1}{2} \hat{\G}_{IJ} \hat{g}^{\mu \nu}
	\partial_\mu \vp^I \partial_\nu \vp^J ,
\end{equation}
where $\hat{\G}_{IJ}(\vp)$ is the Jordan frame field-space metric.
It has been shown that this action can be rewritten in the
Einstein frame after a conformal transformation~\cite{Kaiser:2010ps}
\begin{equation}
	g_{\mu\nu} = f \hat{g}_{\mu\nu} .
\end{equation}
The Einstein frame action is
\begin{equation}
	\label{eq:einstein-action}
	S = \int \d^4 x \, \sqrt{-g} \left[
		\frac{1}{2} \Mpl^2 R
		+ X - V
	\right] ,
\end{equation}
where
$V = \hat{V}/f^2$ is the potential and
\begin{equation}
	\label{eq:X}
	X = - \frac{1}{2} \G_{IJ} g^{\mu \nu} \partial_\mu \vp^I \partial_\nu \vp^J
	, \qquad
	\G_{IJ} = \frac{1}{f} \hat{\G}_{IJ}
		+ \frac{3}{2} \Mpl^2 \frac{f_{,I} f_{,J}}{f^2} .
\end{equation}
Cosmological observables themselves are not altered by this
procedure~\cite{Kaiser:1995nv,Flanagan:2004bz,Chiba:2008ia,Gong:2011qe}.
However, one must be careful to restrict attention to
clearly defined physical quantities;
in particular, the curvature perturbation $\zeta$ is not
itself an observable~\cite{White:2012ya}.

The Einstein-frame action~\eqref{eq:einstein-action} is
of the
same form as~Eq.~\eqref{eq:action}.
Therefore the results
of~\S\S\ref{sec:n-pfs}--\ref{sec:evolution}
are also applicable to models with
non-minimal coupling.

\section{Conclusions} \label{sec:conclusion}

In this paper we have computed
the covariant three-point function near horizon-crossing
for a collection of slowly-rolling scalar fields
with a nontrivial field-space metric.
After making a conformal transformation,
this framework is sufficiently general to include
scenarios where one or more fields are
nonmimimally coupled to the Ricci scalar.
The subsequent superhorizon evolution can be expressed
using a version of the `separate universe' approach.

We concentrate on the broad class of models
described by a $\sigma$-model Lagrangian $\mathcal{L} = X + V$,
where $2X = - \G_{IJ} \partial_\mu \vp^I \partial^\mu \vp^J$,
and obtain expressions for the two- and three-point functions.
The presence of a nontrivial field-space metric leads to
technical subtleties. First,
to obtain a covariant formalism
we must be careful to define perturbations as tangent-space 
vectors $Q^I$ using the method of Gong~\&~Tanaka.
When computing correlation functions involving the $Q^I$,
the Feynman diagram expansion introduces
explicit factors of the `trajectory propagator'
which implements parallel transport along the inflationary trajectory.

Second, new interaction vertices appear which involve
explicit factors of the Riemann curvature tensor.
Therefore
the two- and three-point functions receive
modifications of two types.
The first follow from promotion of the flat field-space
perturbation $\delta \vp^I$ to $Q^I$ and 
covariantize the result 
for $\G^{IJ} = \delta^{IJ}$.
As in general relativity, covariantization is achieved by
exchanging partial derivatives for covariant derivatives
and contracting all indices with the field-space metric.
The second involve explicit factors of the field-space Riemann curvature.
These modify both quantum interference effects operating near
horizon exit, and the interactions between growing modes far
outside the horizon.

In~\S\ref{sec:evolution} we developed a covariant
version of the `separate universe' formalism
to account for these superhorizon-scale interactions.
The covariant Jacobi equation~\eqref{eq:jacobi-eq}
automatically incorporates curvature contributions
which influence the evolution of the two- and three-point function.
We have shown that this 
correctly reproduces the
time-dependent growing modes near horizon-crossing
generated by the apparatus of quantum field theory.
In particular, it matches the
two lowest-order divergences at
single-logarithm order
and the leading divergence
at double-logarithm order.

The Jacobi approach
leads to
covariant `time evolution operators'
${\bigamma^I}_m$ and ${\bigamma^I}_{(mn)}$ 
which can be obtained straightforwardly
by direct integration of~\eqref{eq:gamma2-eq}--\eqref{eq:gamma3-eq}.
Together with the covariant gauge transformations derived
in~\S\ref{sec:observables}
these yield covariant `$\delta N$ coefficients' 
$\biN_m$ and $\biN_{(mn)}$ which define the `separate universe'
expansion of the curvature perturbation $\zeta$ in
Eq.~\eqref{eq:covariant-deltaN}.
This provides a clear and economical framework enabling
perturbations to be evolved in a slow-roll inflationary model with
nontrivial field-space metric.

We always retain the option to abandon manifest covariance
and work with the coordinate variation $\delta\vp^I$.
The traditional
separate-universe expansion for $\delta\vp^I$ is unchanged by the
presence of a nontrivial field-space metric, and
our predictions for the autocorrelation functions of $\zeta$
cannot vary because $\zeta$ is a field-space scalar.
The advantage of the covariant formulation is one of convenience
and practicality.

The phenomenology of the new Riemann-tensor terms may be interesting.
In a canonical scenario, interactions among superhorizon modes
are suppressed by three powers of $\dot{\phi}/H$,
and are therefore relatively slow.
(Note that one should not regard this suppression as an indication of
how large the bispectrum can become.
Rather, it is an indication of the \emph{timescale} over which it
can evolve.) However, Eq.~\eqref{eq:bispectrum} shows that
curvature-mediated interactions are suppressed
by only a single power of $\dot{\phi}/H$.
In a model where the field-space curvature is $\Or(1)$ these could
lead to much more rapid evolution.
It will be interesting to study these effects in more detail,
and we hope to return to this in future work.

\section{Acknowledgements} 

We would like to thank
Chris Byrnes,
Jinn-Ouk Gong,
James Lidsey, David Mulryne,
Courtney Peterson,
Takahiro Tanaka and Shinji Tsujikawa for valuable discussions.
JE is supported by an STFC studentship.
DS acknowledges support from the Science and Technology Facilities Council
[grant number ST/I000976/1]
and the Leverhulme Trust.

JE and RT would like to thank the hospitality of
the Tokyo University of Science,
where some of this work was carried out.
JE 
also thanks
the Royal Astronomical Society for
their support during this visit. 
This work has benefited from exchange visits
supported by a JSPS and Royal Society bilateral grant.

DS would like to thank the
Instituto de Astrof\'{\i}sica de Canarias
and the organizers and participants at
the ISAPP 2012 summer school on `CMB and High Energy Physics'
for their hospitality.
This material is based upon work supported in part by the
National Science Foundation under Grant No.~1066293 and
the hospitality of the Aspen Center for Physics.
\appendix

\section{$n$-point functions in the in--in formalism}
\label{sec:3pfcalcs}
The general theory
necessary to compute $n$-point functions was set out in
the papers by Maldacena~\cite{Maldacena:2002vr}
and Weinberg~\cite{Weinberg:2005vy},
and has been reviewed
elsewhere~\cite{Seery:2007we,Chen:2010xka,Koyama:2010xj}.
In curved field-space this is complicated by the necessity to
ensure that all quantities satisfy the correct tensor
transformation laws.
This means that the structure of the two- and higher $n$-point
functions becomes more elaborate.
Special factors, given by line integrals of the
connection over field space,
are required to ensure their indices reside in the
proper tangent space.

In this Appendix we briefly review necessary elements from the
general theory, focusing on the changes necessary due to
field-space curvature.

\subsection{Two-point function}
\label{sec:appendix-2pf}

After transforming to conformal time $\eta$, defined by
$\d t = a \, \d \eta$,
the in--in
generating functional can be written to quadratic order,
\begin{equation}
	Z = \int [ \d Q_+^I \, \d Q_-^I ]
	\; \exp \left(
		- \frac{\im}{2} \int_{\tau_0}^{\tau} \d^3 x \, \d \eta \; a^2
		\bar{\vect{Q}}^{I}
		\left( \begin{array}{cc}
			\triangle \\
			& -\triangle
		\end{array} \right)_{IJ}
		\vect{Q}^J
		+ \text{$\delta$-fn terms}
	\right) ,
	\label{eq:generating-function}
\end{equation}
where
$\bar{\vect{Q}}^{I} = ( Q^I_+, Q^I_- )$ and
an overbar
denotes matrix transposition.
The time $\tau_0$ should be set
well before horizon crossing of the modes
under discussion, and $\tau$ is the time at which we wish to
compute each correlation function.
The $\delta$-function terms have support at $\tau_0$ and $\tau$,
and enforce boundary conditions, to be described below.
Finally, the differential operator $\triangle$
satisfies
\begin{equation}
	\triangle_{IJ} = \G_{IJ} \left( \Deta^2 + 2 \frac{a'}{a} \Deta
	- \partial_i \partial_i
	\right)
	+ a^2 \M_{IJ} ,
	\label{eq:triangle-operator}
\end{equation}
where $\M_{IJ}$ is the mass matrix~\eqref{eq:mass}.

We write the time-ordered two-point function between contravariant
components of $Q_+$ as $D_{++}$,
\begin{equation}
	D_{++}^{JK'}(\eta, \vect{x} ; \sigma, \vect{y}) =
	\langle \timeorder
		Q_+^J(\eta, \vect{x})
		Q_+^{K'}(\sigma, \vect{y})
	\rangle ,
\end{equation}
with analogous definitions for $D_{-+}$, $D_{+-}$ and $D_{--}$.
In this appendix, we use unprimed indices to label the tangent space at
time $\eta$ and primed indices to label the tangent space at
time $\sigma$.
Each two-point function is a field-space bitensor and a spacetime
biscalar.
The operator $\timeorder$ denotes time ordering, with the convention that
all `$-$' fields are taken later than all `$+$' fields
and that time ordering
on the `$-$' contour is in the reverse sense to the `$+$' contour.

The rules of Gaussian integration enable us to calculate the
$D_{\pm\pm}$. They are obtained by inverting the quadratic
structure $\triangle_{IJ}$,
\begin{equation}
	\im a^2 \left( \begin{array}{cc}
		\triangle \\
		& -\triangle
	\end{array} \right)_{IJ}
	\left( \begin{array}{cc}
		D_{++} & D_{+-} \\
		D_{-+} & D_{--}
	\end{array} \right)^{JK'}
	=
	\G^{K'}_I
	\left(
		\begin{array}{cc}
			1 & 0 \\
			0 & 1
		\end{array}
	\right)
	\delta(\eta-\sigma) \delta(\vect{x}-\vect{y}) .
	\label{eq:propagator-eqn}
\end{equation}
Since an expectation value $\langle \mathcal{O} \rangle$
inherits the tensor transformation properties of $\mathcal{O}$,
the associated two-point functions between covariant or
mixed components of $Q_{\pm}$ can be obtained by raising or lowering
the $J$ and $K'$ indices.
To do so, one should use the metric evaluated at $\eta$
or $\sigma$, respectively.
In~\eqref{eq:propagator-eqn} we have suppressed
coordinate labels, but the differential operator
acts only on $\eta$ and $\vect{x}$.

\paragraph{Mass matrix.}
For the remainder of this section, we ignore the mass matrix $\M_{IJ}$
and treat each mode as massless.
Small masses can be accommodated perturbatively if desired, but we
will not do so in this paper.
Note that this does not imply that we ignore all couplings between
modes \emph{after} horizon crossing: these are certainly important,
because they describe how power is transferred from isocurvature
perturbations to the adiabatic mode.
These couplings will be retained when we discuss time evolution
in \S\ref{sec:evolution}.
We are ignoring them only for a brief period around horizon
exit.

\paragraph{Tensor structure.}
First, consider $D_{++}$.
The $\vect{x}$ and $\vect{y}$ dependence can be diagonalized
by passing to Fourier space,
\begin{equation}
	D_{++}^{JK'} = \int \frac{\d^3 k}{(2\pi)^3}\; D^{JK'}_{++}(\vect{k})
	\; \e{\im \vect{k} \cdot (\vect{x} - \vect{y})} .
\end{equation}
Neglecting the mass matrix as explained above,
the mode function 
$D_{++}^{JK'}(\vect{k})$
satisfies
\begin{equation}
	\G_{IJ} \left( \Deta^2 + 2 \frac{a'}{a} \Deta + k^2 \right)
	D_{++}^{JK'} = - \frac{\im}{a^2} \G_I^{K'} \delta(\eta - \sigma) .
	\label{eq:mode-eqn}
\end{equation}
To factorize the tensor structure we
introduce a bitensor ${\trajp^I}_{K'}$ which is required to
solve the equation
$\Deta {\trajp^I}_{K'} = 0$,
\begin{equation}
	\Deta {\trajp^I}_{K'}
	= \frac{\d {\trajp^I}_{K'}}{\d \eta}
	+ \Gamma^I_{MN} \frac{\d\phi^M}{\d\eta} {\trajp^N}_{K'} = 0 .
\end{equation}
The solution can formally be written as an ordered exponential,
\begin{equation}
	{\trajp^I}_{K'}
	= \pathorder
		\exp \bigg(
			-\int_{\sigma}^{\eta} \d \tau
			\;
			{\Gamma^{I''}}_{M''N''} \frac{\d \phi^{M''}}{\d \tau}
		\bigg)
		\G^{N'}_{K'} ,
	\label{eq:pi}
\end{equation}
where the integral is computed along the inflationary
trajectory and the symbol `$\pathorder$' denotes path ordering
along it.
Eq.~\eqref{eq:pi} is the \emph{trajectory propagator}
introduced in~\S\S\ref{sec:second_order}--\ref{sec:third_order}.
It is simply the parallel propagator evaluated on the
inflationary trajectory.
In~\eqref{eq:pi} we have chosen boundary conditions so that
when $\eta \rightarrow \sigma$ the trajectory propagator
satisfies ${\trajp^I}_{K'} \rightarrow \G^{I}_{K}$.

The trajectory propagator allows the index structure
in~\eqref{eq:mode-eqn} to be factorized. Taking
$D_{++}^{JK'} = \trajp^{JK'} \Delta_{++}$
(where indices on ${\trajp^I}_{K'}$ are raised and lowered
using the usual rules for a bitensor), it follows that
the scalar factor $\Delta_{++}$
satisfies the same equation as the propagator
in flat field-space \cite{Seery:2005gb},
\begin{equation}
	\left(
		\Deta^2 + 2 \frac{a'}{a} \Deta + k^2
	\right)
	\Delta_{++}
	= 
	- \frac{\im}{a^2} \delta(\eta - \sigma) .
\end{equation}
The same factorization can be made for each Green's function,
so $D^{JK'}_{\pm\pm} = \trajp^{JK'} \Delta_{\pm\pm}$.
With vacuum boundary conditions,
the $\delta$-function terms
at $\tau_0$ in~\eqref{eq:generating-function}
require $\Delta_{++}$ to be approximately positive frequency there,
and $\Delta_{-+}$ to be approximately negative frequency.
The $\delta$-function terms
at $\tau$ require $\Delta_{++}(\tau,\sigma) =
\Delta_{-+}(\tau,\sigma)$
for all $\sigma$.
Finally, $\Delta_{--}$ and $\Delta_{+-}$ are the Hermitian conjugates
of $\Delta_{++}$ and $\Delta_{-+}$, respectively.

If the initial conditions
at $\tau_0$ correspond to the vacuum,
then $D_{++}$ should be approximately positive frequency at that time.
If $\tau_0$ is well before horizon-exit, then
\begin{equation}	
	\langle
		\timeorder
		Q_+^I(\vect{k}_1, \eta)
		Q_+^{J'}(\vect{k}_2, \sigma)
	\rangle
	\simeq
	(2\pi)^3 \delta(\vect{k}_1 + \vect{k}_2)
	\trajp^{IJ'}
	\frac{H_\ast^2}{2k^3}
	\times
	\left\{ \begin{array}{l@{\hspace{5mm}}l}
		(1 - \im k \eta) (1 + \im k \sigma) \e{\im k (\eta - \sigma)} &
		\eta < \sigma \\
		(1 + \im k \eta) (1 - \im k \sigma) \e{\im k (\sigma - \eta)} &
		\sigma < \eta
	\end{array} \right. ,
	\label{eq:2pf-pp}
\end{equation}
where $k$ is the common value of $|\vect{k}_1|$ and $|\vect{k}_2|$,
and this estimate is valid for values of $\eta$ and $\sigma$
within a few e-folds of horizon-crossing at
$|k\eta| = |k\sigma| = 1$.
A subscript `$\ast$' denotes evaluation precisely at the horizon-crossing
time.
In Eq.~\eqref{eq:2pf-pp} the trajectory propagator plays an essential
role in ensuring that the right-hand side has the correct
bitensorial transformation law.

Now consider $D_{-+}$.
Imposing the boundary condition that $D_{-+}$ is approximately negative
frequency at the initial time, and equals
$D_{++}$ at time $\tau$, we find
\begin{equation}
	\langle
		\timeorder
		Q_-^I(\vect{k}_1, \eta)
		Q_+^{J'}(\vect{k}_2, \sigma)
	\rangle
	=
	(2\pi)^3 \delta(\vect{k}_1 + \vect{k}_2)
	\trajp^{IJ'}
	\frac{H_*^2}{2k^3}
	(1 + \im k \eta)(1 - \im k \sigma) \e{\im k(\sigma - \eta)} .
	\label{eq:2pf-mp}
\end{equation}

\subsection{Three-point function}
\label{sec:3pf}

Each term in the third-order action~\eqref{eq:s3sr}
makes a contribution to the three-point correlation
function
$\langle Q^I_{\vect{k}_1} Q^J_{\vect{k}_2} Q^K_{\vect{k}_3} \rangle$,
or equivalently $A^{ijk}$.
Vertices are constructed from two copies of the action,
one for each of the `$+$' fields and `$-$' fields of
Eq.~\eqref{eq:generating-function}.
Vertices for `$+$' fields appear with a factor of $+\im$;
vertices for `$-$' fields appear with a factor of $-\im$.
We apply Wick's theorem to produce all ways of pairing
indices, using $G^{++}$ of~\eqref{eq:2pf-pp} to pair
two `$+$' indices and its complex conjugate $G^{--}$ to pair
two `$-$' indices.
We use $G^{-+}$ in~\eqref{eq:2pf-mp} or its complex conjugate
$G^{+-}$ to pair a mix of `$+$' and `$-$' indices.
Finally, we integrate over all possible spacetime positions for
each vertex.

\paragraph{Trajectory propagator.}
In curved field-space, this procedure is modified by the
appearance of the trajectory propagator in each two-point function.
Consider a typical interaction appearing the the
third-order action,
such as the term
\begin{equation}
	S_{(3)} \supseteq
		\int \d^3 x \, \d \eta \; \Big\{
			-\frac{a^2}{4\Mpl^2 H(\eta)}
			\dot{\phi}^{I'} Q_{I'} \Deta Q^{J'} \Deta Q_{J'}
		\Big\} .
\end{equation}
To proceed we should expand background quantities
around a reference scale $k_\ast$, in the direction of the
inflationary trajectory.
This can be determined by analogy with the Taylor
expansion \eqref{eq:covariant-expansion},
which was constructed along a geodesic.
To separate the various tangent-space indices which appear, we adopt
the following conventions:
tangent-space indices at the time of observation,
$\tau$, are labelled $I$, $J$, $K$.
(In the main text, the time of observation is expressed as an
e-folding number $N$.)
Tangent-space indices at the time of horizon-crossing for the
reference scale $k_\ast$
are labelled $i$, $j$, $k$.
(In the main text, the time of horizon-crossing is expressed
as $N_\ast$.)
Finally, indices associated with the integration variable $\eta$ are given
primed indices $I'$, $J'$, $K'$.
The background factor $\dot{\phi}/H$ gives
\begin{equation}
	\frac{\dot{\phi}^{I'}}{H(\eta)}
	=
	{\trajp^{I'}}_i
	\bigg(
		\frac{\dot{\phi}^i}{H}
		+
		\DN \frac{\dot{\phi}^i}{H}
		N
		+ \cdots
	\bigg)_\ast ,
	\label{eq:dotphi-taylor}
\end{equation}
where $N = - \ln |k_\ast \eta|$ represents the number of e-folds
since horizon exit of the reference scale.
We should make an analogous expansion for the metric which is used
to contract the two copies of $\Deta Q^{J'}$,
\begin{equation}
	\G^{J'K'}
	= {\trajp^{J'}}_j {\trajp^{K'}}_k \G^{jk}_\ast .
	\label{eq:metric-taylor}
\end{equation}
In this case, higher-order terms in the Taylor expansion vanish
because the metric is covariantly constant.
A final example is the Riemann tensor, whose series expansion will
be required to compute the higher-order contribution~\eqref{eq:3pfv6},
\begin{equation}
\label{eq:riemann_3pf_expansion}
	R^{I'(J'K')L'} \frac{\dot{\phi}_{L'}}{H}
	=
	{\trajp^{I'}}_i
	{\trajp^{J'}}_j
	{\trajp^{K'}}_k
	\bigg(
		R^{i(jk)l} \frac{\dot{\phi}_l}{H}
		+ \DN
			R^{i(jk)l} \frac{\dot{\phi}_l}{H}
			N
		+
		\cdots
	\bigg)_\ast .
\end{equation}

One can see that the
three factors of the trajectory propagator in
Eq.~\eqref{eq:riemann_3pf_expansion} will be common to all terms
in the three-point function. These propagators carry dependence 
on the integration variable $\eta$ 
as well as the reference time $N_*$. However, each two-point 
function which connects an external
field with a field at the vertex will introduce a
propagator factor ${\trajp^I}_{I'}$, which
is a function of $\eta$ and $N$.
Therefore all $\eta$ dependence cancels, leaving a propagator
that relates the two tangent spaces at $N$ and $N_*$.
This may be factored out of the vertex integral.
The remaining details of the calculation correspond to those
in flat field-space.

\paragraph{Results.}
Three of the contributing terms are covariantized 
versions of those present in flat field-space:
\begin{itemize}
\item $ \displaystyle{-\frac{a^2}{4 \Mpl^2 H} \dot \phi^I Q_I 
\Deta Q^J \Deta Q_J} $ \\
\begin{equation}
	\label{eq:3pfv1}
	A^{ijk} \supseteq  
	-\frac{1}{2 \Mpl^2} \frac{\dot \phi^i_*}{H_*} \G^{jk}_* k_2^2 k_3^2 
	\Bigg( \frac{1}{k_t} + \frac{k_1}{k_t^2}\Bigg) + \text{cyclic} ,
\end{equation}
\item $ \displaystyle{-\frac{a^2}{4 \Mpl^2 H} \dot \phi^I Q_I \partial_i Q^J \partial_i Q_J} $ \\
\begin{equation}
	\label{eq:3pfv2}
	A^{ijk} \supseteq  
	\frac{1}{2 \Mpl^2} \frac{\dot \phi^i_*}{H_*} \G^{jk}_*
	(\vect{k}_2 \cdot \vect{k}_3)
	\Bigg( k_t - \frac{\kappa^2}{k_t} - \frac{k_1 k_2 k_3}{k_t^2} \Bigg)
 	+ \text{cyclic} ,
\end{equation}
\item $ \displaystyle{\frac{a^2}{2 \Mpl^2 H} \dot \phi^I \partial_i \partial^{-2} \Deta Q_I \partial_i Q^J \Deta Q_J} $ \\
\begin{equation}
	\label{eq:3pfv3}
	A^{ijk} \supseteq  
	\frac{1}{2\Mpl^2} \frac{\dot \phi^i_*}{H_*} \G^{jk}_* \Bigg[
	(\vect{k}_1 \cdot \vect{k}_2) k_3^2
	\Bigg( \frac{1}{k_t} + \frac{k_2}{k_t^2} \Bigg)
	+(\vect{k}_1 \cdot \vect{k}_3) k_2^2
	\Bigg( \frac{1}{k_t} + \frac{k_3}{k_t^2} \Bigg)
	\Bigg] + \text{cyclic} .
\end{equation}
\end{itemize}
These expressions are valid under the same conditions as
the power spectrum~\eqref{eq:power-spectrum}:
we must wait sufficiently many e-folds that decaying power-law
terms have become negligible, but Eqs.~\eqref{eq:3pfv1}--\eqref{eq:3pfv3}
are unreliable when $N = -\ln |k_\ast \tau| \gg 1$.
These three contributions can be combined
to give
\begin{equation}
	A^{ijk} \supseteq 
	\frac{1}{\Mpl^2} \frac{\dot \phi^i_*}{H_*} \G^{jk}_* 
	\Bigg[
		-2 \frac{k_2^2 k_3^2}{k_t}
		+ \frac{1}{2} k_1 (\vect{k}_2 \cdot \vect{k}_3)
	\Bigg] + \text{cyclic} .
\end{equation}
We now consider the remaining two terms involving the Riemann tensor:
\begin{itemize}
\item $ \displaystyle{\frac{2 a^3}{3} R^{I(JK)L}\dot \phi_L \Deta Q_I Q_J Q_K} $ \\
\begin{equation}
	\label{eq:3pfv4}
	A^{ijk} \supseteq
	\frac{4}{3} R^{i(jk)m}_* \frac{\dot \phi_m^*}{H_*} 
	\Bigg[
		k_1^3 \bigg( \EulerGamma -N + \ln \frac{k_t}{k_*} \bigg)
		- k_t k_1^2 + \frac{k_1^2 k_2 k_3}{k_t}
	\Bigg] + \text{cyclic} ,
\end{equation}
\item $ \displaystyle{\frac{a^4}{6} R^{(I|LM|J;K)}\dot \phi_L \dot \phi_M Q_I Q_J Q_K} $ \\
\begin{equation}
	\label{eq:3pfv5}
	A^{ijk} \supseteq  
	\frac{1}{3} R^{(i|mn|j;k)}_\ast
		\frac{\dot{\phi}_m^\ast}{H_\ast}
		\frac{\dot{\phi}_n^\ast}{H_\ast}
		\left[
			k_1^3 \Big(
				N - \ln \frac{k_t}{k_\ast} - \EulerGamma - \frac{1}{3}
			\Big) 
			+ \frac{4}{9} k_t^3
			- k_t \kappa^2
		\right] + \text{cyclic} .
\end{equation}
\end{itemize}
Finally we consider the next-order
correction to the
$R^{I(JK)L} \dot \phi_L$ term:
\begin{itemize}
\item $ \displaystyle{\frac{2 a^3}{3H} R^{I(JK)L;M}\dot \phi_L \dot \phi_M N \Deta Q_I Q_J Q_K} $ \\
\begin{equation}
\begin{split}
	\label{eq:3pfv6}
	A^{ijk} \supseteq 
	\frac{2}{3} R^{i(jk)m;n}_* \frac{\dot \phi_m^*}{H_*}
	\frac{\dot \phi_n^*}{H_*} 
	\Bigg[ &
		-k_1^3 N^2 + k_1^3 \bigg(
			\EulerGamma^2 - \frac{\pi^2}{12}
			+ \ln \frac{k_t}{k_*} \Big(
				2\EulerGamma + \ln \frac{k_t}{k_*}
			\Big) \bigg)
	\\ & \mbox{}
		- 2 k_t k_1^2 \Big(
			\ln \frac{k_t}{k_*} + \EulerGamma - 1
		\Big)
		+ 2\frac{k_1^2 k_2 k_3}{k_t}\Big(
			\ln \frac{k_t}{k_*} + \EulerGamma
		\Big)
	\Bigg]
	\\ & \mbox{} + \text{cyclic} .
\end{split}
\end{equation}
\end{itemize}
The presence of
terms proportional to positive powers of $N$, which spoil the
validity of these formulae when $N \gg 1$,
is explicit in Eqs.~\eqref{eq:3pfv4}--\eqref{eq:3pfv6}.

\renewcommand{\baselinestretch}{1}
\bibliography{citations}{}
\bibliographystyle{JHEPmodplain}

\end{document}